%% file: main.tex
\documentclass[conference,compsoc]{IEEEtran}

\makeatletter
\def\ps@IEEEtitlepagestyle{%
  \def\@oddhead{\mycopyrightnotice}%
}
\def\mycopyrightnotice{%
  \begin{minipage}{\textwidth}
  \scriptsize
    Copyright \copyright~2024 IEEE. Personal use of this material is permitted. Permission from IEEE must be obtained for all other uses, in any current or future media, including reprinting/republishing this material for advertising or promotional purposes, creating new collective works, for resale or redistribution to servers or lists, or reuse of any copyrighted component of this work in other works. Full citation: P. Nanayakkara, et al., ``Measure-Observe-Remeasure: An Interactive Paradigm for Differentially-Private Exploratory Analysis,'' in 2024 IEEE Symposium on Security and Privacy (SP), San Francisco, CA, USA, 2024 pp. 231-231. doi: 10.1109/SP54263.2024.00182
  \end{minipage}
}
\makeatother

\usepackage{graphicx}
\usepackage{filecontents}
\usepackage[normalem]{ulem} 
\usepackage{amsmath} 
\usepackage{amsfonts} 
\usepackage{bbm} 
\usepackage{multirow} 
\usepackage{tabularx} 
\usepackage{booktabs} 
\usepackage{mdframed} 
\usepackage{wrapfig} 
\usepackage{floatflt}
\usepackage{floatrow}
\usepackage{subcaption}
\usepackage{hyperref}

\ifCLASSOPTIONcompsoc
  \usepackage[nocompress]{cite}
\else
  \usepackage{cite}
\fi
\ifCLASSINFOpdf
\else
\fi

\begin{document}
%
\title{Measure-Observe-Remeasure:\\ An Interactive Paradigm for Differentially-Private Exploratory Analysis}

\makeatletter
\newcommand{\linebreakand}{%
  \end{@IEEEauthorhalign}
  \hfill\mbox{}\par
  \mbox{}\hfill\begin{@IEEEauthorhalign}
}
\makeatother



%

\author{\IEEEauthorblockN{Priyanka Nanayakkara\IEEEauthorrefmark{1},
Hyeok Kim\IEEEauthorrefmark{1},
Yifan Wu\IEEEauthorrefmark{1},\\
Ali Sarvghad\IEEEauthorrefmark{2},
Narges Mahyar\IEEEauthorrefmark{2},
Gerome Miklau\IEEEauthorrefmark{2},
Jessica Hullman\IEEEauthorrefmark{1}}
\IEEEauthorblockA{\IEEEauthorrefmark{1}Northwestern University}
\IEEEauthorblockA{\IEEEauthorrefmark{2}University of Massachusetts Amherst\\
\{priyankan, hyeokkim2024, yifan.wu\}@u.northwestern.edu\\ \{asarv, nmahyar, miklau\}@cs.umass.edu\\ jhullman@northwestern.edu}}


\input{macros}

\maketitle

\begin{abstract}
Differential privacy (DP) has the potential to enable privacy-preserving analysis on sensitive data, but requires analysts to judiciously spend a limited ``privacy loss budget'' $\epsilon$ across queries. Analysts conducting exploratory analyses do not, however, know all queries in advance and seldom have DP expertise. Thus, they are limited in their ability to specify $\epsilon$ allotments across queries prior to an analysis. To support analysts in spending $\epsilon$ efficiently, we propose a new interactive analysis paradigm, \workflow{}, where analysts ``measure'' the database with a limited amount of $\epsilon$, observe estimates and their errors, and remeasure with more $\epsilon$ as needed.

We instantiate the paradigm in an interactive visualization interface which allows analysts to spend increasing amounts of $\epsilon$ under a total budget. To observe how analysts interact with the \workflow{} paradigm via the interface, we conduct a user study that compares the utility of $\epsilon$ allocations and findings from sensitive data participants make to the allocations and findings expected of a rational agent who faces the same decision task. We find that participants are able to use the workflow relatively successfully, including using budget allocation strategies that maximize over half of the available utility stemming from $\epsilon$ allocation. Their loss in performance relative to a rational agent appears to be driven more by their inability to access information and report it than to allocate $\epsilon$.
\end{abstract}


%
\IEEEpeerreviewmaketitle

\input{01_intro}
\input{02_background}
\input{03_related}
\input{04_interface}
\input{05_evaluation}
\input{06_results}

\input{07_discussion}
\input{08_conclusion_and_acks}

\bibliographystyle{IEEEtran}
\bibliography{ref.bib}

\input{09_appendix}
\input{10_metareview}

\end{document}

%% file: macros.tex
\newcommand{\workflow}[1]{\textsc{Measure-Observe-Remeasure}}
\newcommand{\acs}[1]{\textsc{Census}}
\newcommand{\dia}[1]{\textsc{Diabetes}}
\newcommand{\stu}[1]{\textsc{Student}}

\newcommand{\RPosteriorExAnte}[1]{\textsc{RPosterior}$_\text{Ex-ante}$}

\newcommand{\RPosteriorRand}[1]{\textsc{RPosterior}$_\text{Rand}$}

\newcommand{\RPosteriorZero}[1]{\textsc{RPosterior}$_\text{Zero}$}

\newcommand{\UpperBound}[1]{\textsc{UpperBound}}

\newcommand{\RPrior}[1]{\textsc{RPrior}}

\newcommand{\RPosteriorBehavioral}[1]{\textsc{RPosterior}$_\text{Same}$}

\newcommand{\LowerBound}[1]{\textsc{LowerBound}}

\newcommand{\ReportLoss}[1]{\textbf{reporting loss}}

\definecolor{figYellow}{cmyk}{0,0.5,1,0}
\definecolor{figBlue}{cmyk}{0.75,0.35,0,0}
\definecolor{figGreen}{cmyk}{0.77,0,1,0}
\definecolor{figPurple}{cmyk}{.38,.51,0,0.36}
\definecolor{figPink}{cmyk}{0,1,0.8,0}

\newcommand{\reportLoss}[1]{\textbf{\textcolor{figYellow}{#1}}}
\newcommand{\allocLossOverall}[1]{\textbf{\textcolor{figBlue}{#1}}}
\newcommand{\totalLoss}[1]{\textbf{\textcolor{figGreen}{#1}}}
\newcommand{\allocLossSepAll}[1]{\textbf{\textcolor{figPurple}{#1}}}
\newcommand{\allocLossSepFullBudget}[1]{\textbf{\textcolor{figPink}{#1}}}

%% file: 01_intro.tex
\section{Introduction}
Datasets about people often contain information that is sensitive, but useful to learn in aggregate. For example, published census tables pose the risk of exposing individuals' demographic information, but are necessary for allocating political representation and government funding~\cite{steed2022policy}. Fortunately, approaches based on differential privacy (DP)~\cite{dwork2006calibrating, dwork2014algorithmic} make privacy-preserving analyses on sensitive data possible.

Differentially-private algorithms, however, constrain and reconfigure the data analysis process, posing new challenges for data analysts~\cite{sarathy2023don}. Specifically, differentially-private algorithms inject statistical noise into the analysis process such that increased noise implies stronger privacy guarantees, but lower accuracy of estimates. The amount of expected noise added is controlled by a privacy loss bound defined by $\epsilon$ (the ``privacy loss budget'').

Each time the dataset is queried, the amount of spent $\epsilon$ accumulates. Once $\epsilon$ reaches the maximum bound, an analyst is prevented from issuing further queries. Conducting a differentially-private analysis therefore requires careful considerations around how much $\epsilon$ to spend and on which queries. Thus, it is natural to pre-specify all queries in advance of an analysis so that the mechanism and distribution of $\epsilon$ can be optimized for the query set---for example, by minimizing repetition in information queried from the database---to maximize accuracy. In fact, DP research and real-world implementations tend to fall under the ``query-response'' model~\cite{wasserman2012minimaxity}, which assumes analysts specify all queries in advance and is common in computer science.

However, while this model is naturally supported by DP, it is at odds with the data-dependent process of \emph{exploratory} data analysis (EDA), which is recognized as an integral part of statistical modeling~\cite{tukey1977exploratory}. During EDA, analysts determine subsequent queries based on results earlier in the analysis, meaning they have only a myopic view of future queries and their relative importance. The iterative nature of EDA poses a particular challenge to spending a total privacy loss budget efficiently: for example, suppose an analyst issues a query with an initial amount of $\epsilon$, then later realizes they actually need better accuracy and re-issues the query with a larger amount of $\epsilon$. In this case, the initial amount of $\epsilon$ essentially goes to waste because it still counts toward the total budget but does not contribute to the final query estimate. Picture this pattern over multiple queries, and it is easy to see how analysts---especially those without DP expertise---can quickly burn through their total budget before meeting analysis goals.

Hence, capitalizing on DP's potential for enabling privacy-preserving data analysis requires a new analysis paradigm that balances control and flexibility to support analysts in spending $\epsilon$ efficiently, without an overwhelming number of choices. While full interactivity---where the analyst determines which queries to submit and at which amounts of $\epsilon$---is theoretically appealing, it would require analysts to perform a complex optimization problem with limited information (i.e., the set of future queries) on top of the reasoning that is already entailed in analyzing data. Expecting analysts to solve this problem well does not account for our expectations about human information processing as boundedly rational~\cite{simon1990bounded}. As such, full interactivity could lead analysts to prematurely exhaust their allocated budget or obtain estimates too inaccurate to be useful~\cite{nunez2020every}.

Therefore, we propose the \workflow{} paradigm (i.e., workflow) which helps analysts spend only what they need on each query to accomplish their analysis goals. The paradigm is interactive, such that analysts spend incrementally more $\epsilon$ as they observe estimates and their errors, thus improving estimates at each step (in expectation). First, the analyst makes a query, which is \textsc{Measured} under DP using an initial fraction of the total privacy loss budget. Second, the analyst \textsc{Observes} the estimate. Third, based on the analyst's observations in the second step, they decide whether to \textsc{Remeasure} the query, spending more $\epsilon$ to get additional information about the data, therefore improving the estimate. This process is interactive, such that once the analyst issues a query, they may engage in the \textsc{Observe}-\textsc{Remeasure} loop until the budget is depleted. They may issue and remeasure queries in any order. Further, the workflow employs a mechanism that makes more efficient use of knowledge about the query set by taking into account previous queries and, upon remeasurement, weighting previous estimates with a fresh estimate in a way that yields lower expected error. Remeasurement allows analysts to spend increasingly more $\epsilon$ on a given query, thus avoiding wasting any amount of $\epsilon$.

We (1) instantiate the \workflow{} workflow in an interactive visualization interface that displays query estimates and supports remeasurement. Through a user study, we (2) explore how analysts respond to the paradigm. To analyze results, we (3) extend a rational agent framework for visualization studies with benchmarks designed specifically to evaluate $\epsilon$ allocation, which allows us to (4) investigate opportunities for improvement by comparing analysts' responses to optimal benchmarks under different assumptions. \looseness=-1

Our results indicate that participants' allocation strategies maximize over half the utility that stems from $\epsilon$ allocation. Their loss in performance relative to a rational agent appears to be driven more by their inability to access information and report it than to allocate $\epsilon$. We compare participants' performance to a variety of benchmarks representing different amounts of information and assumptions about allocation strategies to gain insight into how the workflow might be improved in future work.

%% file: 02_background.tex
\section{Background} 

\subsection{Differential Privacy}
Differential Privacy (DP) provides a mathematical framework for accounting for privacy loss incurred during data analysis. In particular, a differentially-private analysis places a limit on how much information about an individual is learned from an analysis based on their data's inclusion in said analysis.

A randomized mechanism \textit{M} satisfies ($\epsilon$)-DP if for any two neighboring datasets \textit{D} and \textit{D'}, which differ by the addition or deletion of one record, and for any subset \textit{S} of the range of \textit{M}, the following inequality holds:

\begin{equation}
Pr[M(D)\in S]\leq exp(\epsilon) \times Pr[M(D')\in S]
\end{equation}

\noindent where $\epsilon$ is a non-negative parameter that controls the strength of privacy protection. DP can be achieved by injecting calibrated statistical noise into numerical query results. $\epsilon$ (i.e., the privacy loss budget) controls the amount of noise the mechanism adds during the computation. A smaller privacy loss budget enforces stronger privacy protection, but typically implies worse accuracy of estimates.

\subsection{Answering and Updating Query Answers}
To build the interface instantiating the \workflow{} workflow we use existing open-source DP methods to answer queries privately, combine noisy queries into consistent estimates, and quantify error. 

\subsubsection{The High Dimensional Matrix Mechanism}
\label{sec:background:hdmm}
Our instantiation of the \workflow{} workflow relies on the High Dimensional Matrix Mechanism (HDMM)~\cite{mckenna2018optimizing, mckenna2021hdmm} to answer queries under DP. HDMM is an extension of the Matrix Mechanism~\cite{li2015matrix} and a state-of-the-art mechanism for answering sets of multi-dimensional linear counting queries. Linear counting queries are a class of queries that include one- and multi-dimensional histograms, marginals, data cubes, etc. 

HDMM takes as input a workload consisting of a set of linear queries. The method computes differentially-private answers to the workload queries by using the Laplace Mechanism~\cite{dwork2014algorithmic, dwork2006calibrating} to answer a different set of queries, called the strategy, from which the workload query answers are derived. It can often increase accuracy (for a fixed value of $\epsilon$) to compute answers in this way. For example, multiple queries in a workload may rely on a common piece of information about the database; HDMM computes a near-optimal set of strategy queries which avoids redundancy and inefficient use of the privacy loss budget. 

In addition to answering queries with reduced error, we use two other features of HDMM to support our interface. First, HDMM is a data-independent mechanism, which allows the straightforward calculation of expected root mean squared error (RMSE) of the reported query answers without spending additional privacy loss budget. We use these values to plot error in our interface. Second, the machinery of HDMM, called inference in Li~et~al.~(2015)~\cite{li2015matrix}, can be used to combine multiple noisy estimates of query answers (in our case derived through remeasures) into a consistent set of estimates with reduced error. We use this feature to combine observations from multiple remeasures and update visualizations shown to the analyst.

%% file: 03_related.tex
\section{Related Work}

Prior work has demonstrated challenges associated with integrating DP into existing workflows, particularly with regard to EDA. In a recent study, Sarathy et al.~\cite{sarathy2023don} interviewed practitioners without DP expertise about their experiences using DP Creator, a DP interface, and found that among other challenges, analysts were skeptical of using DP for EDA given the additional constraint of the privacy loss budget. Through a usability study of four Python-based DP tools, Ngong et al.~\cite{ngong2023evaluating} found that ``novices and experts in [their] study were concerned with setting and tracking the privacy budget.'' While the privacy loss budget represents an unavoidable constraint around how much information can be learned about a dataset, our work seeks to address this challenge by reducing the cognitive overhead associated with keeping track of and spending $\epsilon$ well. Garrido et al.~\cite{garrido2023lessons} interviewed practitioners about hurdles to using DP in enterprise settings. One of their key findings was that DP has potential to ``facilitate exploration that otherwise might not be possible or timely,'' further emphasizing the need to support EDA under DP.

In terms of tool development, prior work has focused on developing interfaces that allow users to input data, queries, and other parameters (e.g., $\epsilon$) and receive query estimates under DP. These tools, described below, make it possible for curators and analysts without DP expertise to employ differentially-private mechanisms. They largely fall under the query response model~\cite{wasserman2012minimaxity}, where there is limited interactivity in terms of spending $\epsilon$. In other words, these tools enact a \textsc{Measure-Observe} paradigm, wherein a user specifies a query to \textsc{Measure} and some amount of $\epsilon$ (or acceptable accuracy or risk) and \textsc{Observes} the result.

Gaboardi et al.~\cite{gaboardi2018psi} introduce PSI ($\psi$) (a pre-cursor to DP Creator), a text-based interface that supports analysts in allocating $\epsilon$ by specifying acceptable amounts of error. Once the analyst finalizes a distribution of $\epsilon$ across queries, they can submit the request (to the curator) and receive results. Thaker et al.~\cite{thaker2020overlook}'s Overlook is a system that supports both data curators and analysts with visualization interfaces for navigating differentially-private analyses. Data curators set a privacy loss budget with which to release a ``synopsis'' of the data to analysts, which analysts can query without limit. Nanayakkara et al. propose ViP~\cite{nanayakkara2022visualizing}, an interface which visualizes trade-offs between accuracy and disclosure risk to help data curators set and split privacy loss budgets. ViP allows curators to test different amounts of $\epsilon$ for a given query, but does not account for privacy loss associated with such testing. While ViP is aimed at a data curator who is permitted to see the raw data, our work is geared toward analysts who do not have such access. St. John et al. introduce the DPP tool~\cite{john2021decision}, which supports a data curator in setting privacy loss budgets by interacting with visualizations depicting risk, sensitivity (of damaged caused by a breach), trust in the recipient, and accuracy. As an alternate approach to helping analysts spend $\epsilon$, Ge et al.~\cite{ge2019apex} propose APEx, a system which allows the analyst to specify error tolerances and in turn receive query results that satisfy said tolerances.\looseness=-1

Finally, researchers have contributed tools and frameworks that support decisions outside setting privacy loss budgets specifically in data analysis contexts. For example, Bittner et al.~\cite{bittner2020understanding} and Guo et al.~\cite{guo2023seeing} contribute interfaces for DP in machine learning contexts. Furthermore, Hay et al.'s DPComp~\cite{hay2016exploring} is a visualization interface that supports data analysts and curators alike in making comparisons between the accuracy of multiple differentially-private algorithms, while Hay et al.~\cite{hay2016principled}'s DPBench formulates an evaluation framework for differentially-private algorithms, where one of the evaluation principles includes an algorithm's inputs, like privacy loss budget. Finally, Zhang et al.'s $\varepsilon$ktelo~\cite{zhang2018ektelo} supports people in designing custom differentially-private algorithms and McSherry's~\cite{mcsherry2009privacy} PINQ is a querying platform for differentially-private analyses.

%% file: 04_interface.tex
\section{\workflow{}}

\begin{figure}[t!]
  \centering
  \includegraphics[width=.8\linewidth]{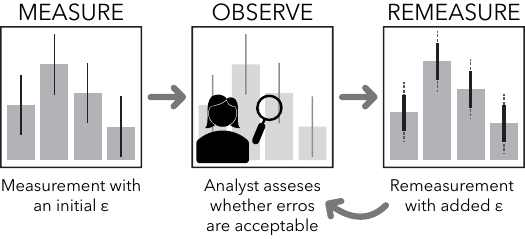}
  \caption{\footnotesize The \workflow{} workflow. The analyst queries the private database for an initial measurement, observes estimates, and remeasures if better accuracy is required.}
  \label{fig:workflow}
\end{figure}

We present the \workflow{} paradigm~(\autoref{fig:workflow}) for EDA under DP. The \workflow{} paradigm assumes an analyst who wishes to explore a private database. They are given a total privacy loss budget---specified by a data curator\footnote{Typically, data curators provide analysts with a total privacy loss budget. However, the paradigm is easily adapted to the setting where the analyst determines the total budget themselves. In such a case, the analyst would set the total budget and proceed in the same way as if the curator had set the total budget.}---to spend during the analysis. Once the analyst depletes the budget, they may no longer issue queries to the database. The analyst's goal is to maximize the accuracy of estimates in such a way that maximizes their utility for inference or decision making. While they \emph{may} begin with some high-level analysis goals (e.g., to identify factors associated with some outcome), they do not know specific queries in advance. We designed the \workflow{} paradigm to be widely accessible, such that analysts without deep knowledge of DP can interactively spend $\epsilon$ without having to get the exact budget allocation for queries ``right'' on the first try.

The analyst issues queries (\textsc{Measure}) and observes (\textsc{Observe}) their estimates, which include DP noise. At any point in their session, if they desire a more accurate estimate for a query, they can \textsc{Remeasure} the query. Remeasuring a query consumes additional $\epsilon$, always yielding lower expected error in the query estimate. The information contained in previous queries is leveraged via the Matrix Mechanism and consequently maximizes the accuracy that can be obtained under the privacy loss budget spent. By using remeasures, they engage in a \workflow{} feedback loop, whereby they continually assess which queries require improved accuracy and distribute remeasures accordingly. In this way, they are not forced to allocate their privacy loss budget up front and can instead spend $\epsilon$ as needed.

The workflow enables analysts to leverage their domain knowledge about the relative importance of each query. For example, an analyst might issue a query to check whether there are roughly an equal number of people in each racial group represented in the dataset. Here, they probably do not require extremely high accuracy, since they are not interested in specific counts, but rather an overall distribution. They may later be interested in learning whether the number of people living in any particular zip code in the dataset exceeds some threshold value. In this case, they may require higher accuracy to be (more) certain whether any counts exceed the threshold.

\begin{figure*}
  \centering
  \includegraphics[width=\textwidth]{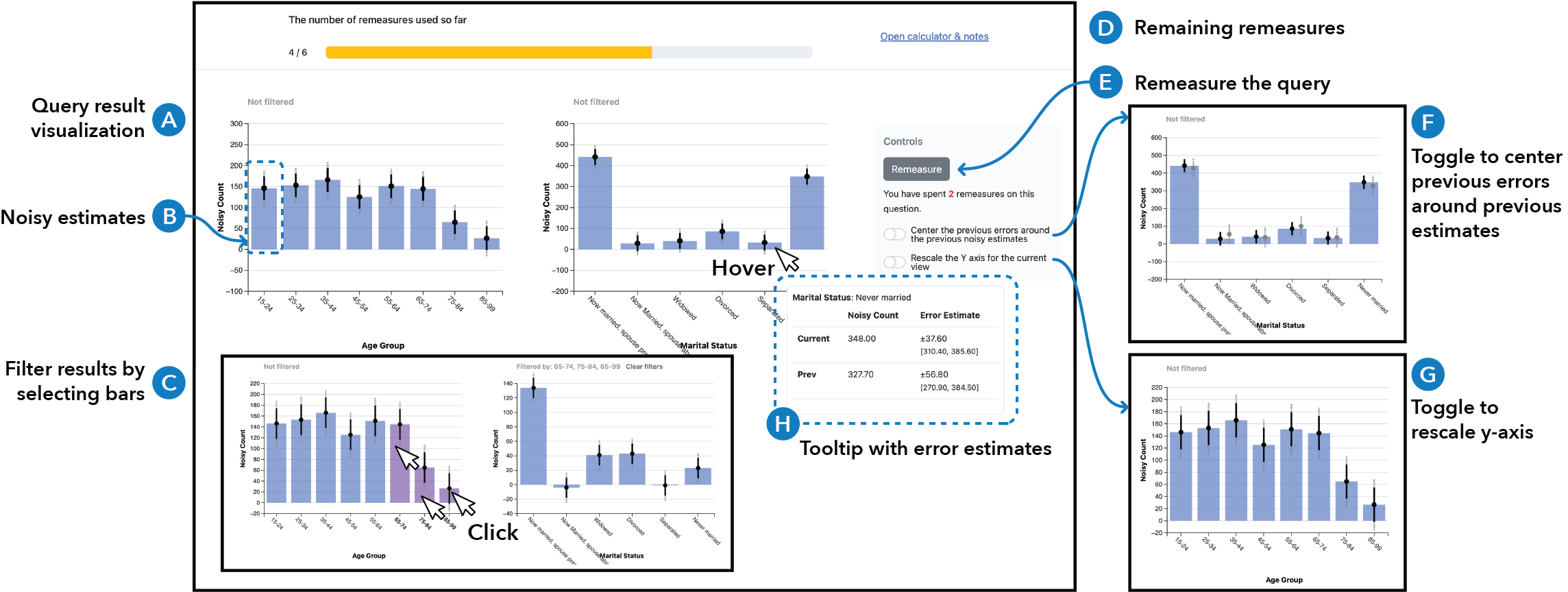}
  \caption{\footnotesize An interactive visualization interface that allows analysts to engage with the \workflow{} workflow. Analysts may observe noisy estimates (\textbf{B}) and remeasure (\textbf{E}) as needed until they reach the total remeasure (privacy loss) budget (\textbf{D}). \looseness=-1}
  \label{fig:interface}
\end{figure*}

\subsection{Implementing Remeasurement} 
In the \workflow{} paradigm, the current set of queries is represented as a workload under HDMM (see Section~\ref{sec:background:hdmm}), for which a strategy matrix is calculated. Query results are computed using the strategy matrix then converted to the response space of the original queries and can then be presented to the analyst. When a query is remeasured, the approach applies a predetermined additional amount of the privacy loss budget to requerying the data, and re-weights new results with all previously cached results (resulting from the original measurement of the data and all consequent remeasurements) to produce an estimate with strictly lower expected error than errors associated with measurements in the cache. We specifically rely on the inference step from HDMM to combine multiple measurements into a single consistent set of estimated query answers. New measurements are weighted as the inverse of their variance.\looseness=-1

The amount of $\epsilon$ applied during each remeasure can be predetermined for the analyst (e.g., by the curator) or set by the analyst. A fixed amount of $\epsilon$ can be applied to each remeasurement or the analyst can adjust exactly how much $\epsilon$ is applied on each remeasurement. However, this level of flexibility adds complexity. Fixing $\epsilon$ per remeasurement helps minimize how much added cognitive complexity DP introduces into analysts' typical EDA workflows.

\subsection{Operationalizing the Paradigm in an\\Interactive Visualization Interface}
\label{sec:workflow:interface}

In theory, analysts with deep DP expertise could implement the machinery of the \workflow{} paradigm as described above and perform exploratory analyses within the paradigm. However, making the paradigm usable by analysts without DP expertise warrants operationalizing it such that they are able to make key decisions (i.e., how to allocate remeasures) without being exposed to the underlying DP machinery. An interface for doing so can take several forms, from a Python or R package with functions for using the Matrix Mechanism to measure and remeasure queries to a visualization-based graphical user interface. In this work, we focus on the latter to explore the potential for the workflow to support budgeting decisions when users lack much DP expertise.

We operationalize the \workflow{} paradigm through an interactive visualization interface\footnote{\url{https://interactive-dp-analysis.github.io/}} (\autoref{fig:interface}). The goal of the interface is to support analysts without deep DP expertise in (1) remeasuring and tracking accumulated privacy loss budget (in the form of remeasures), (2) comparing accuracy of estimates across queries at any given time, and (3) observing changes in error upon remeasurement. In our instantiation, we assume the data curator has pre-specified the amount of $\epsilon$ per remeasure, but the interface can be easily extended to allow analysts flexibility to set budget-per-remeasure.

\textbf{Visualizations.} The interface visualizes all queries issued at any given point on a scrollable page. This enables the analyst to compare relative errors across issued queries to decide how to allocate remeasures.\footnote{Note that initial measurements support the query set, but might not be the optimal measurements to answer all queries.} For each query, the interface displays one visualization per each variable specified in the query (\autoref{fig:interface}\textbf{A}). For example, a query about conditional counts of people in a dataset according to age and marital status would be displayed in two visualizations: (1) a histogram depicting noisy counts by age group and (2) a bar graph depicting noisy counts by marital status. These visualizations are displayed side-by-side and are ``linked'' in such a way that the analyst can filter data displayed on either visualization by making specifications on the other visualization (\autoref{fig:interface}\textbf{C}). For example, if the analyst is interested in the distribution of marital-status groups across particular age groups, they can click the bars associated with the age groups of interest, and the marital status visualization will only display data for the specified age groups. \looseness=-1

Points at the top of each bar represent noisy estimates as computed by HDMM (\autoref{fig:interface}\textbf{B}). The interface also displays error bars representing RMSE of the estimates. Upon hovering a bar, a tooltip appears with values of the noisy count, error estimate, and an interval representing the noisy count $\pm$ error. Visually displaying error bars allows analysts to gauge how precise an estimate is, and in turn assess whether they want to allocate more $\epsilon$ to a given query.

\textbf{Remeasuring.} The analyst may remeasure any of the displayed queries by using remeasure buttons in each query's control panel~(\autoref{fig:interface}\textbf{E}), to the right of its visualizations. Clicking the remeasure button remeasures the query with the fixed amount of $\epsilon$ allocated to each remeasure. The analyst is guaranteed a reduction in the expected error each time a remeasure is applied to a query. A progress bar at the top of the screen shows the number of remeasures used across all queries, supporting the analyst in keeping track of how much of the total privacy loss budget they have spent~(\autoref{fig:interface}\textbf{D}). The progress bar remains fixed at the top of the page as the analyst scrolls through visualizations of multiple queries' estimates. Each query's control panel also keeps track of the number of remeasures used on that particular query. 

When an analyst remeasures a query, the visualization for that query updates to show the new noisy estimates and error estimates. The error estimates from the measurement directly preceding are also shown as dotted gray bars beneath the new error bars; the previous errors are also centered around the current estimate (\autoref{fig:interface}\textbf{B}). Displaying both sets of error bars, centered around the same point, enables analysts to easily compare widths of previous and current errors to assess the improvement in accuracy between measurements.

\textbf{Adjusting Visualizations.} If the analyst wants to see estimates from the immediately-preceding measurement, they can use the first toggle in the control panel for the query. Toggling shifts the dashed-line previous errors slightly to the right and centers them around previous estimates (\autoref{fig:interface}\textbf{F}). A second toggle in each query's control panel automatically re-scales the visualizations' y-axes such that the maximum limit better reflects the largest estimate displayed (\autoref{fig:interface}\textbf{G}). Re-scaling can be useful when the analyst has filtered the displayed data on a visualization and wants to ``zoom in'' by reducing the axis limit. Similarly, it can be useful when a remeasure results in a large decrease in estimated error and the analyst similarly needs to zoom in. By default, this toggle is off so that upon remeasuring, analysts see updated estimates on a common scale. \looseness=-1

%% file: 05_evaluation.tex
\section{Observing the Paradigm: User Study}
The primary difference between the \workflow{} paradigm and the current practice in differentially-private EDA (i.e., the \textsc{Measure-Observe} paradigm), is the interactivity afforded by \workflow{}. Under the current paradigm, analysts are forced to make high-stakes guesses about how much $\epsilon$ to apply to a query: if they spend some amount of $\epsilon$ and decide it was not enough, and consequently re-query with higher $\epsilon$, the initial amount of $\epsilon$ does not contribute to the final estimate, yet counts toward their total budget use. Thus, there is inherent performance loss due to wasted $\epsilon$ (i.e., not being able to spend $\epsilon$ interactively). The amount of loss is a function of several factors, including the dataset, query, and how performance is scored (i.e., the utility function). To provide intuition of how much performance gain is possible using our paradigm over the \textsc{Measure-Observe} paradigm, we provide a comparison of RMSE of estimates under both paradigms across multiple datasets and queries from analysis tasks in our user study, in Appendix~\ref{sec:evaluation:MO_vs_MOR}.

In practice, however, there are various sources of human error that could impact how effectively people use the \workflow{} paradigm. Therefore, we conducted an exploratory user study to observe possible variations in performance that could occur in practice, and gain insight into reasons why analysts may struggle with the paradigm as implemented in the visualization interface. For example, analysts may differ in how well they allocate $\epsilon$ across queries to maximize utility or interpret noisy estimates to form beliefs about the true (un-noised) data. Specifically, we study the following:

\begin{enumerate}
    \item How well the interface supports analysts in using noised query results to accurately answer analysis questions
    \item How well the interface supports analysts in making efficient use of a total privacy loss budget, through remeasurement, to complete analysis tasks
    \item Which aspects of the interface are perceived as challenging versus helpful, and high-level appraisals of the \workflow{} workflow broadly
\end{enumerate}

Participants conducted analyses by using differentially-private query results to answer questions about multiple datasets, with the option of using some total budget of remeasures per dataset. In addition to payment for completing the study, participants also received bonus payments commensurate with the accuracy of their responses. We formalize the decision problem they faced and compare the payoffs they received to those expected under different assumptions about their decision strategy, including random allocation of remeasures and optimal reallocation as defined in a rational agent framework~\cite{wu2023rational}.

\subsection{Empirical Evaluation Setting}
There is flexibility in decisions an analyst can make when using the workflow (e.g., which queries to issue and how much $\epsilon$ to spend per remeasure). However, it is difficult to compare performance across analysts or to a notion of best possible performance without a well-defined decision problem. We designed an empirical setting in which participants used the \workflow{} workflow to answer specific questions about a common set of datasets using a limited number of remeasures. We defined proper scoring rules that dictated the utility associated with better answers to a query in order to ensure that each analyst was equally incentivized to maximize their score and consequently, the bonus payment they earned during the study. Thus, their goal was to provide answers to the questions (and therefore allocate the remeasures) that maximized their payoff under the given scoring rules.

Each participant in our experiment completed three blocks of questions and stimuli. Each block posed four analysis questions about a particular dataset, to be answered with up to six remeasures (each of $\epsilon = 0.3$). Thus, each block contained four visualizations (each contained a pair of linked histograms---see Section~\ref{sec:workflow:interface}) each corresponding to a different analysis question. Participants were not restricted in how they could allocate the remeasures across the visualizations (queries).

\subsubsection{Eliciting and Scoring Responses}
Because we presented \emph{estimates} (i.e., DP-noised query results), we expected participants to be uncertain about the ground truth values. Thus, we elicited answers that reflected the uncertainty in their beliefs. We asked quantitative and binary questions (further described in Section~\ref{sec:user_study:question_types}), where quantitative questions asked for a count and binary questions asked whether a given count was above or below a certain threshold. For quantitative questions, participants provided an interval in which they were 95\% confident the true value fell. For binary questions, participants provided the probability they would assign to the ground truth being ``yes'' and the probability they would assign to the ground truth being ``no,'' where the two probabilities must sum to one. \looseness=-1

A scoring rule evaluates the quality of a probabilistic forecast by assigning a numerical score based on the prediction and the event or the actual value~\cite{gneiting2007strictly}. We scored responses using proper scoring rules, where, conditional on a belief distribution $Q$, the agent cannot do better by reporting some other distribution $P \neq Q$. We used an interval scoring rule~\cite{gneiting2007strictly} for quantitative questions and the Brier (i.e., quadratic) scoring rule for binary questions~\cite{brier1950verification}:

\begin{itemize}
    \item \emph{Interval Scoring Rule:}\\ $ S_\alpha^{\text{int}}(l,u;x) =  u - l + \frac{2}{\alpha}\mathbbm{1}\{x < l\} + \frac{2}{\alpha}(x - u)\mathbbm{1}\{x > u\}$, where $l$ and $u$ are the lower and upper bounds on the participant's reported interval, $x$ is the ground truth, and $\alpha = .05$
    \item \emph{Brier/Quadratic Scoring Rule: } \\
    $S(p, \theta)= (p-\mathbbm{1}(\theta=\text{yes}))^2-((1-p)-\mathbbm{1}(\theta=\text{no}))^2$, where $p$ is the participant's probability guess for ``yes'' and $\theta$ is the ground truth
    
\end{itemize}

The interval scoring rule rewards tighter intervals that contain the true value; that is, there is a penalty if the interval does not contain the true value and the score is worse for wider intervals. The Brier scoring rule rewards probabilities that are closer to the ground truth (e.g., if the ground truth is ``yes,'' the ground truth answer is probability 1 for ``yes'' and probability 0 for ``no'').

In order to associate each question with the same maximum possible payment of \$2.50 (up to \$10 per block), we normalized scores to [0,1] by dividing by an expected maximum bound on each question's score (see Appendix~\ref{appendix:normalizing_scores} for details). Across all three blocks, participants could earn up to a total of \$30 (in addition to a guaranteed payment of \$25 for completing the study).

\subsubsection{Datasets}
Each block asked questions about one of the datasets described below. We chose these datasets because they are inherently about people, thus creating realistic analysis scenarios where DP might be applied. However, we believe showing these datasets to participants posed negligible real-world privacy risks to people in the datasets because these data are not only all available (without identifiers) on the UC Irvine Machine Learning Repository, but participants in our study were also only shown differentially-private estimates of the data.

\smallskip \noindent
\textbf{NIST Diverse Communities Data~\cite{NIST_ACS_data} (\acs{})}: The data are based on the U.S. Census Bureau's American Community Survey, and includes demographics like race, age, and income.

\smallskip\noindent
\textbf{Diabetes 130 Dataset~\cite{strack2014impact} (\dia{})}: The dataset describes patient visits at 130 U.S. hospitals over a ten-year period and includes information like number of medications and length of hospital stay.\looseness=-1

\smallskip\noindent
\textbf{T\"{u}rkiye Student Evaluation Dataset~\cite{GunduzFokoue:2013} (\stu{})}: The data describe student evaluations of courses and instructors at Gazi University. Students answered questions rating various aspects of courses and instructor performance.

In each block, participants were shown a sample of size 1,000 from each of the three datasets. Their responses were scored against the query results for that sample.

\subsubsection{Analysis Question Types}
\label{sec:user_study:question_types}
Each block contained three quantitative questions and one binary question. Quantitative questions asked for the number of people in the dataset satisfying some criteria (e.g., by race and age). Binary questions asked whether the number of people in the dataset satisfying some criteria was lesser or greater than a threshold value. We selected threshold values that were purposefully challenging, such that the actual count was near the threshold and the first error interval participants encountered would likely contain the threshold. Examples of each question type drawn from the \acs{} block are below: \looseness=-1

\smallskip\noindent
\textbf{Quantitative.} How many ``Black or African American alone'' and ``Asian alone'' people are at least 55 years old?

\smallskip\noindent
\textbf{Binary.} Are there more than 327 people who have never been married and make less than \$100,000?

\smallskip
To add further variety and realism in question types, we asked a combination of ``multi-value'' and ``single-value'' questions. Multi-value questions required participants to sum multiple given values, while single-value questions required reading only one value. Both questions above are multi-value questions. For example, answering the quantitative question requires filtering either the race or age visualization and summing the appropriate bars---e.g., one can sum the number of ``Black or African American people alone'' people who are at least 55 years old with the number of ``Asian alone'' people who are at least 55 years old. There were two multi-value and one single-value quantitative questions per block. Two blocks contained a multi-value binary question while one contained a single-value binary question. All questions are in Appendix~\ref{appendix:task_questions}.

\subsubsection{Protocol}
Participants gave informed consent prior to beginning study sessions. We gave participants a high-level introduction to the concept of injecting noise during an analysis, then walked them through a tutorial question to familiarize them with the interface and its features.\footnote{In an initial round of sessions, we mistakenly described error estimates as representing mean squared error (MSE) vs. RMSE (correct). We omit participant data from these sessions, except in two cases where participants said in a follow-up that they either thought error represented RMSE or would not have changed their responses had they been told RMSE.} Participants could ask any questions about the interface and about remeasuring. We then explained the scoring rules. For the duration of the study, they were allowed to reference a document with the scoring rules and an explanation of how scores would be converted to payoffs.\footnote{See here for the scoring rule document, full protocol, and interface shown to participants: \url{https://interactive-dp-analysis.github.io/}} 

Participants then completed each block (four questions, up to six remeasures). We counterbalanced block order and dataset version, and randomized the order of questions within blocks. Participants could complete questions within a block in whatever order they wished. We provided them with a simple calculator, but they were permitted to use their own if they preferred. There was no strict time limit on each block, however we gave them warnings to ensure they completed the study in the allotted session time (one hour). Finally, participants answered a series of exit interview questions (Appendix~\ref{appendix:exit_interview}), including about their remeasure strategy, how much of the \$30 payoff they believed they earned, and their familiarity with DP.

Our study included 14 participants who were based in the U.S., at least 18 years old, and had experience with quantitative data analysis. We recruited participants by posting on listservs for graduate students in computer science or computer-science-adjacent fields and through our networks. We did not recruit directly from any courses taught by the authors. The first author conducted sessions virtually, and upon completion, participants were given gift cards (\$25 for completing the study plus some portion of the \$30 bonus). The study was deemed exempt by Northwestern University's IRB. \looseness=-1

\section{Rational Agent Benchmarks and Losses}
\label{sec:analysis:rational_agent}

We define best attainable performance and other meaningful comparison points in order to evaluate study results. These comparison points allow us to contextualize participants' performance and reflect on sources of observed performance loss.

\subsection{Benchmarks}
\label{sec:benchmarks_losses:benchmarks}
To devise comparison points (i.e., benchmarks) we use a rational agent framework based in statistical decision theory. The framework uses the notion of a rational Bayesian agent to quantify the maximum amount of information that can be learned from some stimuli (i.e., visualization of a noisy estimate) and applied to a decision problem (i.e., forecasting the true value underlying the estimate)~\cite{wu2023rational}. At a high level, we conceive of a rational agent who begins with a prior over possible ground truth answers, is presented with the same study tasks as participants, perfectly perceives the presented stimuli, Bayesian updates their prior, and submits an optimal forecast. The expected performance of the rational agent upper bounds participants' performance. 

To apply the rational agent framework, we first formalize the decision problem induced by our study by defining the payoff-relevant state (i.e., ground truth answers to the queries), a data-generating model (DGM) that defines a joint distribution over signals (noised visualizations that inform of the payoff-relevant state) and the state, and a scoring rule that maps an agent's forecast and the ground truth answer to a payoff. The rational agent's prior is defined by assuming the rational agent has knowledge of the DGM that produces the experimental stimuli. In other words, the rational agent understands how the visualizations are created: they are aware of the ground truth dataset versions, and that for each block, one of four versions is drawn and noised using some known initial $\epsilon$.

We use the framework to devise benchmarks representing worst-case lower and best-case upper bounds on performance under different conditions. Unless otherwise noted, each benchmark quantifies the expected per-block payoff defined over blocks and where applicable, over specific seeds (for noisy estimates) and dataset variations participants received. Specific calculations are available in Appendix~\ref{appendix:benchmarks}. \looseness=-1

We begin by designing a worst-case expected payoff, \LowerBound{}. This benchmark assumes participants form responses \emph{before} seeing any noisy estimates and only using information available to them before beginning the study (the size of each dataset). This benchmark represents a baseline payoff if a participant did not pay any attention to the tasks and answered questions without using any information beyond the task instructions.

\begin{mdframed}
\textbf{\LowerBound{}.} 
The expected payoff of an agent who forms fixed forecasts by only using knowledge of the dataset size (e.g., by evaluating the CDF of a discrete uniform distribution from 0 to 1,000).
\end{mdframed}

\LowerBound{} is not a tight lower bound on the rational agent's payoff since they, unlike participants, know the full experimental design when starting the study. Hence, we design a worst-case lower bound on payoff for the rational agent, \RPrior{}, which is obtained when they use a best fixed strategy to determine forecasts accounting only for their prior knowledge (i.e., they do not account for any noisy estimates). \looseness=-1

\begin{mdframed}
\textbf{\RPrior{}.}
The expected payoff of a rational agent who takes the best fixed action by accounting only for their prior (i.e., by uniformly sampling over ground truth answers and using the appropriate quantiles or proportions of draws to determine responses). 
\end{mdframed}

Next, we define an upper bound, \UpperBound{}, on performance for the rational agent, which naturally also upper bounds participants' performance. Defining \UpperBound{} also allows us to quantify the total gain in payoff possible ($\text{\UpperBound{}} - \text{\LowerBound{}}$). The best possible payoff is obtained when the rational agent optimally allocates remeasures such that information learned from the resulting noisy estimates maximizes their posterior scores. Unfortunately, identifying the optimal allocation is an NP-hard problem, hence computationally infeasible. Thus, to approximate the true upper bound, we instead assume the rational agent spends six remeasures per query. Assuming they spend six remeasures \emph{per query} upper bounds payoffs since the study task only allows six remeasures total \emph{across queries}. This is a tighter upper bound than the simplest \$10 upper bound (total possible payoff per block).

\begin{mdframed}
\textbf{\UpperBound{}.} 
The rational agent's expected payoff when spending six remeasures \emph{per} query.
\end{mdframed}

Next, we define benchmarks between \LowerBound{} and \UpperBound{} that can be used to disambiguate sources of participants' payoff loss, including their remeasure allocation strategies. This benchmark, \RPosteriorExAnte{}, allows us to compare participants' allocations to that of a rational agent who chooses their allocation strategy before seeing any noisy estimates. \looseness=-1

\begin{mdframed}
\textbf{\RPosteriorExAnte{}.}
The rational agent's expected payoff when using the optimal fixed (\emph{ex-ante}) allocation strategy. This strategy is one that maximizes expected payoff over possible dataset versions and possible noisy estimates (admitted by the DP mechanism) without knowledge of any noisy estimates.\looseness=-1
\end{mdframed}

We next define a benchmark to help study whether participants' allocations were better than random, and therefore whether they appear to have obtained useful information from the visualizations:

\begin{mdframed}
\textbf{\RPosteriorRand{}.}
The rational agent's expected payoff when randomly allocating the remeasure budget across queries. Note: we calculate expected per-block payoffs, over many possible random allocations, that the rational agent would earn assuming they see the same noisy estimates and receive the same dataset versions as participants.
\end{mdframed}

The above benchmarks alone do not allow us to directly quantify the value obtained from remeasuring. Thus, we define a benchmark corresponding to the rational agent not spending \emph{any} remeasures, but still seeing the initial set of visualizations (i.e., those that appear on the interface when it first loads). Comparing this benchmark to payoffs under strategies where remeasures are spent characterizes the value attributable to remeasurement.

\begin{mdframed}
\textbf{\RPosteriorZero{}.}
The rational agent's expected payoff when they do not spend any remeasures. In other words, this payoff corresponds to when they Bayesian update on \RPrior{} using only the initial set of measurements. \looseness=-1
\end{mdframed}

Last, we are interested in understanding how much the rational agent would have earned had they used the same remeasure allocation strategies as participants. This benchmark allows us to, for example, isolate payoff loss due to allocation decisions separate from participants being non-Bayesian (because the comparisons we make are in rational agent space, where the agent is Bayesian).

\begin{mdframed}
\textbf{\RPosteriorBehavioral{}.} The rational agent's expected payoff when making the \emph{same remeasure allocations as participants}. \looseness=-1
\end{mdframed}

\subsection{Losses}
\label{sec:benchmarks_losses:losses}
Participants can lose payoff due to two main sources: not optimally allocating remeasures across queries and being non-Bayesian (including not having the experimental design, Bayesian updating, and optimally reporting forecasts based on noisy estimates). The rational agent framework enables us to infer how much payoff loss for participants stems from each source using the benchmarks defined above. We define the following losses in total possible payoff, where $P$ denotes the average payoff participants earned per block:

\begin{itemize}
    \item \textbf{\ReportLoss{}.} The loss in payoff due to participants not reporting the optimal beliefs, conditional on their remeasure allocation, due to not being Bayesian. Unlike participants, the rational agent always Bayesian updates their prior based on noisy estimates, which they perfectly interpret from visualizations. Hence, the difference in payoff when the rational agent and participants use the same remeasure allocation strategy ($\text{\RPosteriorBehavioral{}} - P$) must stem from either participants not having access to the prior/using a different prior, not updating their beliefs like a Bayesian, or not accurately interpreting the noisy estimates. We quantify this loss as the fraction of the total payoff increase theoretically possible that the difference represents ($\frac{\text{\RPosteriorBehavioral{}} - P}{\text{\UpperBound{}} - \text{\LowerBound{}}}$).
    \vspace{1.5mm}

    \item \textbf{allocation loss (overall).} The loss in payoff due to participants suboptimally allocating remeasures across queries. The difference between \UpperBound{} and \RPosteriorBehavioral{} captures only the loss owing to participants' allocations because it is computed in rational agent space, where being Bayesian is held constant. We contextualize this difference by dividing by the total possible payoff increase possible. Thus, we compute this loss as $\frac{\text{\UpperBound{}} - \text{\RPosteriorBehavioral{}}}{\text{\UpperBound{}} - \text{\LowerBound{}}}$.
\end{itemize}

Note that the above losses are on the scale of possible payoff increase, which includes both decisions participants must make, reporting and allocation. We can separately look at the scale for payoff increase from allocation alone: $\text{\UpperBound{}} - \text{\RPosteriorZero{}}$. The baseline $\text{\RPosteriorZero{}}$ corresponds to the rational agent who does not spend any remeasures but perfectly reports information. We additionally compute allocation loss (separated) which captures the fraction of payoff loss under this scale of allocation payoff increase without the confounding factor of reporting loss:

\begin{itemize}
    \item \textbf{allocation loss (separated).} We compute allocation loss (separated) as $\frac{\text{\UpperBound{}} - \text{\RPosteriorBehavioral{}}}{\text{\UpperBound{}} - \text{\RPosteriorZero{}}}$. When we compute allocation loss (separated) filtered on blocks where participants spent the full remeasure budget, the denominator becomes $\text{\UpperBound{}} - \text{\RPosteriorRand{}}$ because \RPosteriorRand{} represents a tighter lower bound on the minimum payoff where all six remeasures are spent.
\end{itemize}

%% file: 06_results.tex
\begin{figure*}[ht]
\RawFloats
	\centering
    \includegraphics[width=\textwidth]{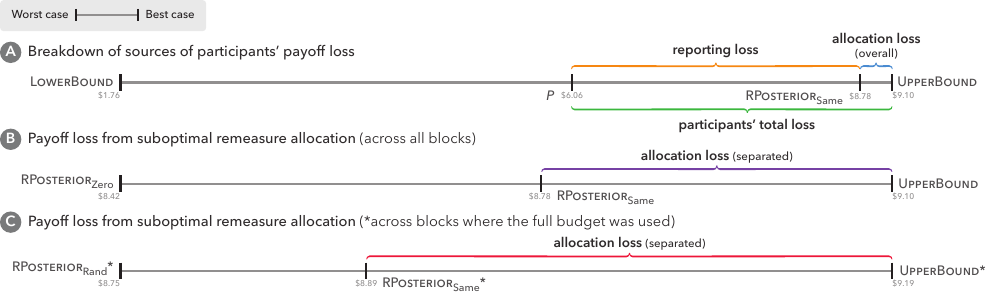}
	\caption{\footnotesize Comparisons between participants' performance and rational agent benchmarks.}
 \label{fig:comparisons}
\end{figure*}

\section{Results}
\label{sec:results}
We begin with preliminaries, including participants' backgrounds, then we present a descriptive analysis of participants' responses, and finally we benchmark their performance within the rational agent framework described above.\footnote{Code to simulate the rational agent and compute benchmarks, as well as data and descriptive analysis code are available here: \url{https://osf.io/ewgkx/?view_only=7fefb3b220ab46178e4002c9235b5a71}}\looseness=-1 

\subsection{Preliminaries}
Participants answered a total of 126 quantitative questions (3 per block $\times$ 3 blocks $\times$ 14 participants) and 42 binary questions (1 per block $\times$ 3 blocks $\times$ 14 participants). 

\textbf{Participants' Backgrounds.} Participants described their level of familiarity with DP as an average of 2.2 ($\text{median}=2$; ``slightly familiar'') on a 5-point Likert scale. Those with at least some familiarity understood the general concept behind DP---for example, by reading an article or watching a video---but were not DP experts. Participants all had experience with quantitative data analysis, and their experiences ranged across domains (e.g., environmental data, health data, social networks). Nine participants had backgrounds leaning more toward classical statistics (vs. predictive modeling or machine learning broadly) while five said their backgrounds leaned more toward predictive modeling.\looseness=-1

\begin{figure}[t]
  \centering
  \includegraphics[width=\linewidth]{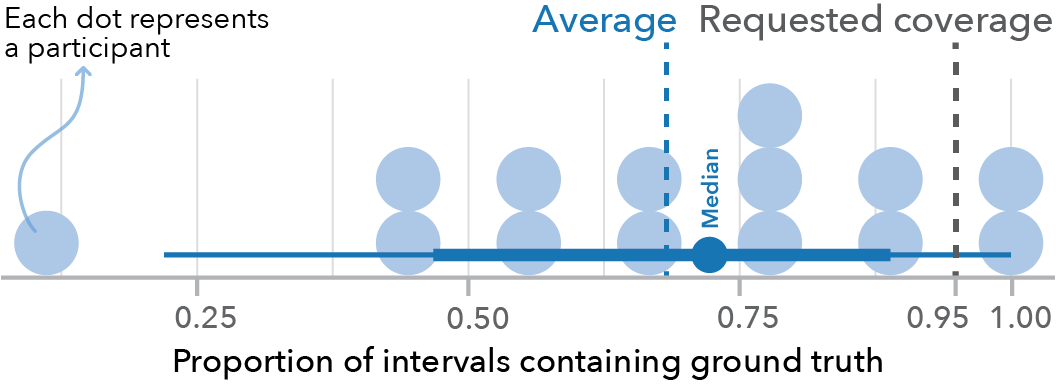}
  \caption{\footnotesize The plot shows the proportion of intervals that contained the ground truth answer, per participant. Each dot represents a participant. The vertical blue line depicts the mean proportion (0.68) and the vertical gray line shows 0.95, the coverage participants were asked to provide.}
\label{fig:prop_intervals_containing_ground_truth}
\end{figure}

\subsection{Descriptive Statistics}
\textbf{Accuracy, Quantitative Questions.} Nearly 70\% of interval responses contained the ground truth answer. Two (out of 14) participants always provided intervals containing the ground truth, while 11 participants provided intervals containing ground truth in over half of their intervals. \autoref{fig:prop_intervals_containing_ground_truth} shows the proportion of intervals containing ground truth, per participant. Most participants provided intervals with empirical coverage less than 95\%, as predicted by prior research on overconfidence in intervals elicited from experts~\cite{soll2004overconfidence}.\looseness=-1

The average width of a participant's interval was about 24 units, while the median was about 18. Recall that the number of records in each block's dataset was 1,000, so ground truth answers could have ranged from 0 to 1,000. Intervals containing ground truth (mean width $\approx$ 29) tended to be wider than intervals not containing ground truth (mean width $\approx$ 13). The proportion of intervals containing ground truth across single- and multi-value questions were similar (0.71 and 0.67, respectively), but the mean interval width for multi-value questions ($\approx 31$) was over triple that of single-value questions ($\approx 9$).\looseness=-1

\textbf{Accuracy, Binary Questions.} Mean absolute error in ``yes'' probability responses (calculated relative to the ground truth of 0 or 1) was 0.37 (95\% CI: [0.26, 0.49]). Over 60\% of probability responses were in the same direction as ground truth---that is, rounding responses to a whole number yields the ground truth. Participants tended to provide ``extreme'' answers, where the probability allocated to ``yes''/``no'' were close to 0 or 1: over 70\% of probabilities assigned to ``yes'' were either less than or equal to 0.2 or greater than or equal to 0.8. As a result, when participants provided answers in the opposite direction to ground truth, they were off by a substantial amount (an average of 0.80). Participants were slightly more likely to answer single-value binary questions in the correct direction than multi-value questions (71\% vs. 57\%). \looseness=-1

\subsubsection{Remeasure Allocation}
Participants allocated remeasures fairly uniformly across questions, spending an average of 1.3 remeasures per question (median $=1$). There were no instances of participants spending more than four remeasures on a single question; in less than 3\% of cases did participants spend more than two remeasures on a question. While participants usually spent all allotted remeasures per block, in over a quarter of completed blocks (11 out of 42), participants spent \textit{fewer} than six remeasures. When asked during exit interviews about why they did not spend all available remeasures, one participant said they forgot they had remeasures in the first block, presumably because they were occupied with the task of answering questions alone. Another participant attributed not using all remeasures to an issue of time management---if they were to remeasure, they said they would want to recalculate their response. Hence, we believe that participants understood that remeasuring improved estimates, but failed to use all remeasures for other reasons. The mean number of remeasures spent per block was about five (median $=6$). Further details on remeasures by question type and participants' distributions of remeasures over questions are in Appendix~\ref{appendix:extra_analysis:payoff_by_qtype}; \autoref{fig:remeasures} in Appendix~\ref{appendix:remeasure_allocations} shows the distribution of participant remeasures (in gray) across blocks.

\subsubsection{Average Earnings and Self-Rated Confidence}
Participants earned an average payoff of \$6.06 per block (out of a total \$10 at the block level). Averages per block follow: \acs{} $=\$5.44$; \dia{} $=\$6.33$; \stu{} $=\$6.40$. They received an average of \$18.17 out of the \$30 bonus available across all three blocks. When asked how much of the bonus they thought they earned, participants said an average of \$13.86. On average, the difference between actual payoffs and perceived payoffs was \$4.31 and only four participants earned less than their perceived payoff. Details on payoffs per question type are in Appendix~\ref{appendix:extra_analysis:payoff_by_qtype}. When asked to rate their confidence in their answers on a scale from 0--100, participants said an average of about 75 (median $\approx 78$).

\subsection{Benchmarking Participants' Performance}
We compare participants' performance to that expected under the benchmarks defined above, and compute losses in payoff owing to errors in reporting and remeasure allocation. \looseness=-1

\looseness=-1

\subsubsection{Amount of \emph{Possible} Payoff Increase Participants Earned}

To begin, we compute the amount of possible payoff increase participants actually earned ($P - \LowerBound{}$) over the total possible payoff increase ($\UpperBound{} - \LowerBound{}$)\footnote{As described in Section~\ref{sec:benchmarks_losses:benchmarks}, \UpperBound{} upper bounds the actual upper bound on total payoff possible. We cannot compute the true upper bound, but know that it lies between \RPosteriorExAnte{} and \UpperBound{}; since the values for both are close (\$8.88 and \$9.10, respectively), we proceed as if \UpperBound{} is the true upper bound.}. We find that participants earned 59\% of the possible increase (i.e., they lost 41\% of the possible increase [\totalLoss{participants' total loss}; \autoref{fig:comparisons}\textbf{A}]), suggesting that although participants earned a substantial amount of the potential increase, there is still room for improvement. Next, we investigate specifically how much payoff increase participants lost due to reporting and allocation decisions.

\subsubsection{Losses in Participants' Payoff Increase}
Recall that participants could lose possible payoff increase due to imperfect reporting (reporting loss) and imperfect remeasure allocations (allocation loss [overall]). We quantify and compare how much of the total possible payoff increase participants lost based on each source of error.

We find that participants lost an average of 37\% of possible payoff increase due to \reportLoss{reporting loss} (\autoref{fig:comparisons}\textbf{A}). Recall that reporting loss accounts for loss owing to participants not being exposed to the DGM, not perfectly perceiving presented information, and not Bayesian updating. Hence, from these results alone, we cannot know whether (1) providing participants with more information about the experimental design (i.e., the prior), (2) helping them more accurately perceive presented information, or (3) helping them Bayesian update would help close the gap, but this is a fruitful area of future work (e.g., through calibration~\cite{wu2023rational}, which helps further disambiguate sources of error).

Next, we investigate payoff loss resulting from participants suboptimally allocating remeasures (i.e., allocation loss [overall]). We find that \allocLossOverall{allocation loss (overall)} (\autoref{fig:comparisons}\textbf{A}) is 4\%; that is, participants lost only 4\% of total possible payoff as a result of their allocation strategies. Thus, participants lost a greater fraction of total payoff increase due to imperfect reporting vs. remeasure allocation.

\subsubsection{Quality of Participants' Allocation Strategies}
We further study participants' remeasure allocation strategies by controlling for reporting loss, thus isolating errors owing to imperfect allocation. We control for reporting loss by setting the baseline as $\text{\RPosteriorZero{}}$ and calculating allocation loss (separated) as defined in Section \ref{sec:benchmarks_losses:losses}.

We first investigate whether participants' allocations appear to be responsive to seeing noisy estimates. We check on a block-by-block basis whether the rational agent who makes the same allocation as the participant (\RPosteriorBehavioral{}) earns more than if the rational agent makes a random allocation (\RPosteriorRand{}). If $\text{\RPosteriorBehavioral{}} > \text{\RPosteriorRand{}}$, we conclude that the allocation used by the participant appears to be responsive to seeing noisy estimates. We find that in just over half of blocks (55\%), participants' remeasure allocations appear better than random.\footnote{For each of the 42 completed blocks, we compare the rational agent's payoff with the same remeasure allocation strategy the participant used with the rational agent's payoff with a random allocation strategy.} The fact that in some blocks participants did not allocate better than random could be due to randomness in strategies (even if participants use a random strategy, there is a chance their payoff is lower than the expected payoff over random allocations).

Now, recall that in a little over a quarter of blocks, participants did not use the full remeasure budget. We find that in blocks where participants used the full remeasure budget, they had a higher chance of \RPosteriorBehavioral{} $>$ \RPosteriorRand{} (65\% vs. 27\%). In fact, excluding blocks where participants did not use the full budget, \RPosteriorBehavioral{} (\$8.88) is close to \RPosteriorExAnte{} (\$8.94), indicating that when using all remeasures, participants made decisions comparable to a strong allocation strategy, albeit one that is determined ahead of seeing noisy estimates.

Next, we turn our attention to allocation loss (separated), which tells us the fraction of \emph{payoff increase possible assuming perfect reporting} that participants' strategies lost. We find that \allocLossSepAll{allocation loss (separated)}(\autoref{fig:comparisons}\textbf{B}) is 47\%, indicating that participants' strategies maximized over half the utility stemming from allocation, but there is still room for improvement. Note that allocation loss (separated) is on a tighter scale that controls for reporting loss, hence the higher value compared to allocation loss (overall). Next, we compute \allocLossSepFullBudget{allocation loss (separated)} (\autoref{fig:comparisons}\textbf{C}) only on blocks where the full remeasure budget was used. Future tools can easily prompt analysts to spend any remaining budget, and therefore it is useful to learn how much loss still occurs once controlling for spending the full budget. On blocks where all remeasures were spent, we find that participants lost 70\% of possible payoff (in this case, \UpperBound{} - \RPosteriorRand{}). Considering we use a strong worst-case payoff (\RPosteriorRand{}) in this calculation, analysts achieving 30\% (i.e., losing 70\%) of the possible payoff increase represents a relatively substantial portion of the possible payoff increase.

\subsection{Exit Interview Findings}

\subsubsection{Remeasure Allocation Strategies}
Participants employed a variety of strategies when deciding how to allocate remeasures. Nine participants (P1-5, P8-9, P11, P13) described using the size of error estimates in some capacity, usually allocating remeasures to visualizations with large error estimates. As P13 described:

\begin{quote}
    I was basically looking at upper bound minus lower bound for whichever questions I feel like [the gap is] a lot. I was trying to get it as tight as possible by using the remeasures.
\end{quote}

P9 adopted this strategy after first trying a uniform allocation strategy:
\begin{quote}
    At first I just spread [remeasures] across [visualizations] fairly equally, but then as I went through the study I realized there were definitely a range of variances across the different questions . . . because we are awarded for having tighter ranges, I started using the remeasures on [visualizations] where there was a wider variation in the data.
\end{quote}

In general, applying remeasures uniformly, at least as some part of the allocation strategy, was not uncommon. This strategy is consistent with our finding that the average number of remeasures spent on each question is slightly over one. Four participants (P4, P6, P9-10) described first spending one remeasure on each question, then using different strategies to determine where to spend remaining remeasures. For example, P4 said they allocated remeasures uniformly, but held off on spending remeasures on visualizations with smaller error bars.

Two participants (P6, P10) looked for ``trends'' in how visualizations changed between remeasures. For example, P6 applied remeasures to visualizations which showed large changes in estimates:

\begin{quote}
 I would look at the first mean and the mean of the first remeasure. If those means are far apart, then I would remeasure.
\end{quote}

A few participants described allocating remeasures in ways that were independent of the error in estimates and scoring rules, suggesting some deviation from the instructions. Three participants described allocating remeasures based on question type. P8 and P5 tended to allocate remeasures to multi-value questions, and P2 prioritized spending remeasures on binary questions. One participant (P14) applied remeasures when error intervals contained negative values and two participants (P11-12) described spending remeasures on either the first or last couple visualizations. P11 said that if they had extra remeasures by the time they answered the last question, they would spend extra remeasures on that visualization.

\subsubsection{Future Strategies}
When asked how they would approach the analysis tasks differently were they to re-do them, four participants (P10, P12-13, P4) said they either would not change their strategy or were unsure about how they would change their approach. On the other hand, five participants (P1-3, P9, P11), two of whom used strategies that were independent of error in estimates, described paying more attention to error estimates in some form, either by interpreting them differently or reviewing error estimates across queries before allocating remeasures. For example, P1 said they would first filter visualizations to focus on estimates of interest, then observe errors across relevant estimates, and complete ``all questions sort of simultaneously.'' Two participants (P6, P8) said they would employ the strategy they developed in later blocks on the first block. These results suggest that use of the \workflow{} workflow may improve as analysts become more familiar with it, and that it might be useful to suggest to analysts in real, evolving EDA workflows that they proceed through queries without remeasuring at first, then circle back to apply remeasures once the query set has been determined.

\subsubsection{Interface Feedback}
We summarize feedback participants gave during exit interviews on their experience using the \workflow{} paradigm via the interface. A handful of participants mentioned remeasuring (P2-3, P7), the graphs generally (P6-7, P14), the tooltip (P6-7, 9), and the ability to recenter errors (P1, P11) as helpful. Notably, participants' independent, positive appraisals of remeasuring suggests that the interactivity of the paradigm was simple for participants to adapt to.

Over half of participants (P2, P5-6, P8-13) mentioned filtering as a helpful feature and exactly half of participants (P2-4, P7, P9, P14) said that the ability to re-scale the y-axis was helpful. These findings suggest that DP tools supporting histogram queries may benefit from enabling filtering via brushing \& linking, and the ability to adjust y-axis limits to customize how zoomed in the view of the data is. Last, participants provided other interface-level feedback that can inform future DP tools. Three participants (P4, P9, P12) wanted the tool to support automatic summation of selected noisy estimates or said that summing presented challenges. Finally, four participants (P2-3, P9-10) did not find re-centering around previous errors to be useful, two (P2, P11) wanted to tooltips to remain ``on'' when hovered (ostensibly so they could reference numbers as they made calculations), and one (P12) wanted to see all previous errors displayed. \looseness=-1

%% file: 07_discussion.tex
\section{Limitations}
Our study results are subject to some limitations. First, we cannot separate loss from not understanding scoring rules from the sources of loss we describe above. Our results may also be sensitive to the elicitation interface we used. Further, due to a bug in our code for generating differentially-private estimates, variable labels within two questions were flipped. Thus, the noisy estimates shown to participants actually corresponded to a different subset of groups than expected. We therefore scored participants based on the ground truth underlying the estimates they saw. During analysis, we recomputed the normalization constants for these two questions using the same process as before, but accounting for the flipped labels. We doubt this had much, if any, impact on results, but we cannot say for sure.

Mismatches between our experimental setting and real-world analysis settings may also impact the validity of our results. We provided participants with datasets and analysis questions, whereas in practice, analysts iteratively form analysis questions on datasets of their choosing during an EDA. However, as an initial step in studying our paradigm, we were interested in comparing analysts' performance to best attainable performance, which is hard to do without a well-defined decision problem. Future work may conduct additional studies where participants define their own analysis questions, perhaps on a dataset of their choosing, and qualitatively evaluate their analyses on the basis of how well analysts felt they met their EDA goals.

Further, our study is exploratory by nature and attempts to get an initial sense of where participants struggled when using the paradigm, and gain insight into why they may have struggled. Our results, which are based on a relatively small sample size, therefore should not be interpreted as confirmatory evidence of analysts' performance under our paradigm. Future work might attempt larger, controlled studies that compare analysts' use of our paradigm to their use of approaches based in the \textsc{Measure-Observe} paradigm.\looseness=-1

\section{Discussion}
\subsection{Making Differential Privacy Usable}
DP has been adopted by several large organizations (e.g., Apple~\cite{apple2017}, Google~\cite{erlingsson2014rappor}, Microsoft~\cite{ding2017collecting}, Uber~\cite{tezapsidis2017uber}, and the U.S. Census Bureau~\cite{abowd2018us}) in recent years, but has yet to be widely adopted by smaller organizations across industry~\cite{garrido2023lessons}. DP still has a high barrier to entry, often requiring expert teams to implement differentially-private systems, therefore making it less likely that smaller organizations can use it. Our work attempts to make it easier for organizations without DP experts to conduct differentially-private analyses by using our approach.

Given the fact that our study participants completed tasks relatively well despite limited training, the \workflow{} paradigm may help enable more organizations to adopt DP. Future work should expand tools that support analysts, for example by creating tools that support the \workflow{} paradigm for other queries common in data analysis (e.g., mean, maximum, minimum). Supporting other queries may benefit from visualization strategies beyond histograms, depending on the nature of the query. For example, displaying differentially-private scatterplots~\cite{zhang2016challenges, zhou2022dpviscreator, panavas2023investigating} raises questions around how to show changes in estimated error upon remeasurement. In such cases, animation or other display strategies may be useful in conveying changes in error~\cite{hullman2015hypothetical}.

Additionally, somewhat surprisingly, participants in our study did not always use the full remeasure (privacy loss) budget, despite being explicitly told that remeasuring improves estimates. Based on observation and probing during exit interviews, we believe that the cognitive load of interpreting noisy estimates and forming beliefs was great enough at times to cause participants to forget about or discount the value of remeasuring. This unexpected behavior highlights the challenge of supporting analysts in adopting DP in their workflows, and suggests that additional support in managing $\epsilon$ could be beneficial. Apart from simple reminder mechanisms to use the full budget, tools might also suggest various allocation strategies for analysts to consider. Analysts can then adjust strategies according to their needs and the estimates they observe.

We envision also supporting analysts in making more efficient use of $\epsilon$ by allowing query results to update ``in tandem'' when $\epsilon$ is spent on any given question. For example, if two questions overlap in data schema, results from one question can potentially increase the accuracy of the other. Thus, when an analyst spends $\epsilon$ on one question, our interface could also dynamically update plots for other questions that can also ``gain'' from the additional information obtained for the first question. HDMM supports such updating. However, if this feature is integrated, analysts may need additional support to determine how to adjust their strategies to account for multiple query revisions in tandem.

\subsection{Adapting the Rational Agent Framework to Differential Privacy}
Our work contributes an adaption of a rational agent framework to DP---specifically to evaluate privacy loss budget decisions against various benchmarks representing different strategies and access to information in a way that allows for identifying and characterizing sources of error. If adapted in the literature on interfaces for DP decision-making, this approach offers a rigorously-defined notion of best attainable performance with an interface. The framework can be applied to improve experiment design before a study is run by identifying where the incentive for participants to use the query results is low relative to the baseline expected payoff for completing the experiment (for example, for our study, this incentive can be calculated as \UpperBound{} $-$ \LowerBound{}). Future work may also involve calibrating participants' responses to account for their not knowing the experimental design, allowing further disambiguation among sources of behavioral error~\cite{wu2023rational}. Calibration entails scoring the expected behavior of a rational agent who has access to the joint distribution of the uncertain state and the participants' reports, but not the signals (i.e., visualization).
Note that the framework can be applied to evaluate tools beyond interfaces, such as DP querying systems or libraries~\cite{garrido2021get}, as long as they support a common set of analysis tasks.

%% file: 08_conclusion_and_acks.tex
\section*{Conclusion}
We propose the \workflow{} paradigm for helping analysts conduct differentially-private exploratory analyses. We instantiate the paradigm in an interactive interface, which allows analysts to interactively spend $\epsilon$ by observing results and allocating additional $\epsilon$ where necessary. In this way, the paradigm aids in staying under a total privacy loss budget while obtaining high-utility results. We explore how analysts interact with the paradigm through a decision-theoretic user study, which allows for comparisons in utility obtained by participants with utility obtained by a rational agent faced with the same decision task. We find that analysts are able to use the paradigm relatively successfully; they are able to maximize over half the utility stemming from allocation of $\epsilon$. Compared to allocating $\epsilon$, they lose more payoff from suboptimally interpreting noisy estimates and using them to form beliefs. \looseness=-1

\ifCLASSOPTIONcompsoc
  \section*{Acknowledgments}
\else
  \section*{Acknowledgment}
\fi

The authors would like to thank members of the MU Collective for feedback on the interface. Gerome Miklau, Narges Mahyar, and Ali Sarvghad were supported by NSF SATC grant \#1954814.

%% file: 09_appendix.tex
\appendices 

\section{Comparing to \textsc{Measure-Observe}}
\label{sec:evaluation:MO_vs_MOR}

We provide some intuition around the benefits of the \workflow{} paradigm by comparing accuracy outcomes between our paradigm and the \textsc{Measure-Observe} paradigm across example datasets and different amounts of $\epsilon$.

Imagine the following scenario: An analyst is conducting an EDA and midway through, they issue a query to learn how many people satisfying a given set of attributes (e.g., race and age) are in the dataset. We assume they use HDMM and the Laplace Mechanism to compute differentially-private estimates. They \textsc{measure} the query with $\epsilon = x$ to get an initial estimate, $p_{original}$. They quickly realize the error associated with the estimate (relative to the estimate itself) is too large, and decide to spend an additional $\epsilon = 2x$ on the query. Depending on which paradigm they are operating under, they spend the additional $\epsilon$ in different ways:\looseness=-1

\smallskip
\noindent \textbf{\textsc{Measure-Observe} Paradigm.} The analyst decides to \textsc{measure} the query again with $\epsilon=2x$, which results in a new estimate $p_{new}$. They \textsc{observe} $p_{new}$ and the associated error and are satisfied. They proceed with the EDA to other queries. Note that the full analysis of this query consumed a total of $\epsilon=3x$, but $p_{new}$ was produced with only $\epsilon=2x$.

\smallskip
\noindent \textbf{\workflow{} Paradigm.} The analyst decides to \textsc{remeasure} the query with $\epsilon = x$. They \textsc{observe} the new estimate, $p_{interim}$, which already accounts for information learned from $p_{original}$ and a fresh estimate. They quickly realize they require even better accuracy, so they \textsc{remeasure} again with $\epsilon=x$. The resulting estimate, $p_{new}$, accounts for information learned from $p_{original}$, $p_{interim}$, and another fresh estimate. They are satisfied with $p_{new}$ and proceed with the EDA. Like in the \textsc{Measure-Observe} version of the scenario, the full analysis of this query consumed a total of $\epsilon=3x$, however in this case, $p_{new}$ was also produced with $\epsilon = 3x$.

We compare errors associated with each paradigm under the above scenario on count queries on three datasets for three example queries (\acs{}, Question 1; \dia{}, Question 1; \stu{}, Question 1)\footnote{For each dataset, we use one of the dataset variations that appeared in the user study.} for different starting values of $\epsilon \in [0.1, 0.3, 0.5]$.

\begin{figure}[h]
  \centering
  \includegraphics[width=\linewidth]{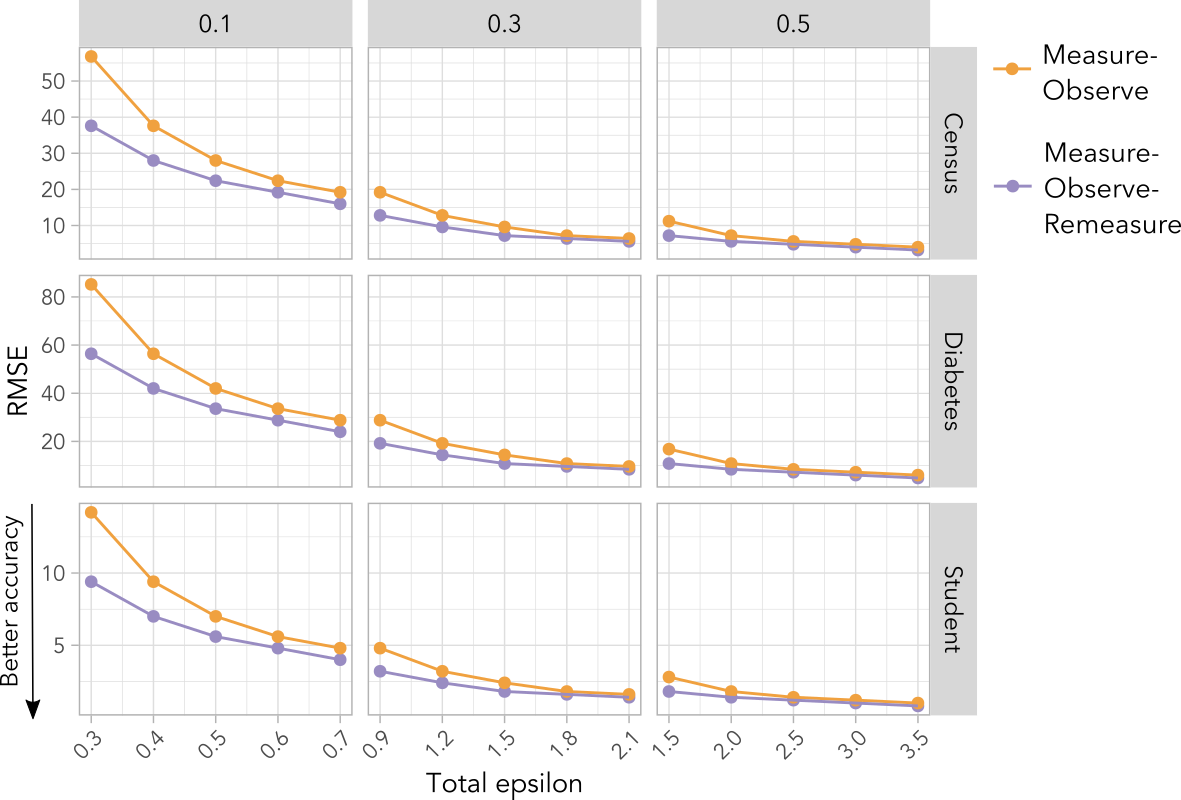}
  \caption{\footnotesize The plot shows differences in expected errors obtained when the analyst continues the analysis under the \textsc{Measure-Observe} vs. \workflow{} paradigms. For the same amount of total $\epsilon$ spent, the \workflow{} paradigm leads to smaller expected error, owing to the fact that all previous estimates are combined with a fresh estimate to produce the new estimate.}
\label{fig:MO_MOR_comparison}
\end{figure}

Figure~\ref{fig:MO_MOR_comparison} shows expected errors of the estimates the analyst will receive under each paradigm. The errors are computed using HDMM's error equation, which is data-independent. The first points on both curves show error obtained from the example scenarios above, where the analyst allocates $\epsilon = 2x$ after realizing the initial amount of $\epsilon = x$ was insufficient, where $x$ is the starting value of $\epsilon$. The following points show scenarios where the analyst decides to allocate more $\epsilon$ to the query after realizing the initial amount was insufficient (i.e., they then instead allocate $\epsilon = 3x$, $\epsilon=4x$, etc.).

\section{Normalizing Scores}
\label{appendix:normalizing_scores}
We normalize scores for a given question using $\frac{x-\text{score}}{x}$, where $x$ is a normalization constant. We use question-specific normalization constants representing a ``maximum bound'' on each question's score. The normalization constant for all binary questions is $2$, since this is the highest (i.e., worst) possible score produced by the quadratic scoring rule. On the other hand, scores produced by the interval scoring rule are unbounded. Thus, to obtain ``maximum bounds,'' for quantitative questions, we imagine that a participant always reports the (noisy) estimate $\pm$ error as their lower and upper bounds, and use the $95^{\text{th}}$ quantile of scores produced with this method as the normalization constant (see \autoref{tab:normalization_constants}). If normalized scores are negative (i.e., because a participant's answer exceeded our prediction of our expectations about bounds), we convert the score to $0$. Participants could earn a maximum of \$2.50 per question, depending on the accuracy of their answers.

\begin{table}[h]
\begin{tabular}{|l|c|c|}
\hline
\textbf{Dataset} & \multicolumn{1}{l|}{\textbf{Question}} & \multicolumn{1}{l|}{\textbf{Normalization Constant}} \\ \hline
\acs{}              & 1                                      & 75.00                                                \\ \cline{2-3} 
                 & 2                                      & 2.00                                                 \\ \cline{2-3} 
                 & 3                                      & 217.69                                               \\ \cline{2-3} 
                 & 4                                      & 56.8                                                 \\ \hline
\dia{}              & 1                                      & 113.4                                                \\ \cline{2-3} 
                 & 2                                      & 283.3                                                \\ \cline{2-3} 
                 & 3                                      & 186.2                                                \\ \cline{2-3} 
                 & 4                                      & 2.00                                                 \\ \hline
\stu{}              & 1                                      & 178.8                                                \\ \cline{2-3} 
                 & 2                                      & 185.6                                                \\ \cline{2-3} 
                 & 3                                      & 2.00                                                 \\ \cline{2-3} 
                 & 4                                      & 56.8                                                 \\ \hline
\end{tabular}
\caption{Normalization constants.}
\label{tab:normalization_constants}
\end{table}

\newpage
\section{Task Questions}
\label{appendix:task_questions}

Below are analysis questions shown to participants. We randomly ordered questions within each block.\\

\acs{}
\begin{enumerate}
    \item How many ``Black or African American alone'' and ``Asian alone'' people are at least 55 years old? (\textit{quantitative; multi-value})
    \item Are there more than 327 people who have never been married and make less than \$100,000? (\textit{binary; multi-value})
    \item How many ``White alone'' people do not have children? (\textit{quantitative; single-value})
    \item How many people who are at least 65 years old are widowed or divorced? (\textit{quantitative; multi-value})
\end{enumerate}

\dia{}
\begin{enumerate}
    \item For people who had elective or newborn admission types, how many had hospital stays for at least 3 days? (\textit{quantitative; multi-value})
    \item How many people under 60 years old take at least 10 medications? (\textit{quantitative; multi-value})
    \item How many African American people had hospital stays lasting less than 5 days? (\textit{quantitative; single-value})
    \item Are there 450 or fewer people who had emergency admission types and are 50 years or older? (\textit{binary; single-value})
\end{enumerate}

\stu{}
\begin{enumerate}
    \item How many students said that they agreed or strongly agreed that Instructor 3’s knowledge was relevant and up-to-date? (\textit{quantitative; single-value})
    \item How many students said that they strongly disagreed or disagreed that they greatly enjoyed and were eager to participate in Instructor 1 or 2’s class? (\textit{quantitative; multi-value})
    \item Are there more than 167 students who rated their attendance as 2 or less, and the difficulty of the course as 4 or more? (\textit{binary; multi-value})
    \item How many students said they agreed or strongly agreed that the instructor’s knowledge was relevant and up-to-date, and also rated the difficulty of the class as 3 or less? (\textit{quantitative; multi-value})
\end{enumerate}

\section{Exit Interview Questions}
\label{appendix:exit_interview}
\begin{enumerate}
    \item Describe your strategy for deciding where to allocate remeasures.
    \item How well do you think you did on the task?
    \item How would you rate your confidence in your answers on a scale from 0 to 100, where 100 is the most confident.
    \item How much of the \$30 bonus do you think you earned?
    \item Briefly describe which parts of the interface you found challenging to use (and didn't seem to help you with the task).
    \item Briefly describe which parts of the interface you found easy to use (and helped you with the task).
    \item What information would you have liked the interface to have included to better support you with the task?
    \item If you were to re-do the task, what would you do differently?
    \item How would you describe your level of familiarity with differential privacy? (1 – Not at all familiar 2 – Slightly familiar 3 – Somewhat familiar 4 – Moderately familiar 5 – Extremely familiar)
    \item Briefly describe the kind of analysis you tend to do or have a background in.
\end{enumerate}

\section{Calculating Benchmarks}
\label{appendix:benchmarks}
\subsection{Calculating \RPrior{}}
We define the rational agent prior responses for each quantitative question by first sampling many draws (e.g., $10,000$) with replacement from the four ground truth answers (corresponding to each dataset version for the block) to mimic the sampling procedure used in the study. The rational agent's interval belief about the ground truth answer is then bounded by the $2.5^{\text{th}}$ and $97.5^{\text{th}}$ quantiles of the draws. For each binary question, we imagine the rational agent samples $10,000$ draws with replacement from the four ground truth answers, and uses the proportion of ``yes''/``no'' answers from the draws as their beliefs about the ground truth answer. We convert these prior responses first into scores, then into expected payoffs. With expected payoff per question, we then calculate expected payoff per block.

\subsection{Calculating \LowerBound{}}
To compute \LowerBound{}, we use only information about the task setup without any noisy estimates: that there were 1,000 records per dataset, each block contained four questions (three quantitative, one binary), and answer formats (intervals, probabilities) for each question type. Hence, for quantitative questions, we assume a discrete uniform distribution over possible counts $x \in \{0, 1, ..., 1000\}$ and compute the $2.5^{\text{th}}$ and $97.5^{\text{th}}$ quantiles of the distribution as interval lower and upper bounds, respectively. Binary questions ask whether the count of records $x$ satisfying a set of constraints are above/below some threshold $t$. If a question asks, for example whether $x$ is less than $t$, we use $\text{Pr}[x < t]$ as the probability of ``yes'' and $1 - \text{Pr}[x < t]$ as the probability of ``no.'' We convert responses into scores and expected payoff per question, then block by summing.

\section{Additional Analysis Results}
\label{appendix:extra_analysis:payoff_by_qtype}

\subsection{Payoffs by Question Type}
We examine differences in payoffs across types of questions---each question was worth \$2.50. Participants earned an average of \$1.41 (median $=\$1.65$) for quantitative questions and an average of \$1.82 (median $=\$2.39$) for binary questions. There appears to be a difference between payoffs for single- and multi-value questions: the average payoff for single-value questions was \$1.97 (median $= \$2.36$) while the average payoff for multi-value questions was \$1.29 (median $= \$1.62$).

\subsection{Remeasure Allocations}
\label{appendix:remeasure_allocations}
Participants spent an average of 1.3 remeasures on quantitative questions and an average of 1.1 remeasures on binary questions. Participants spent an average of about one remeasure on single-value questions and an average of about 1.4 remeasures on multi-value questions.

\begin{figure}[h]
  \centering
  \includegraphics[width=\linewidth]{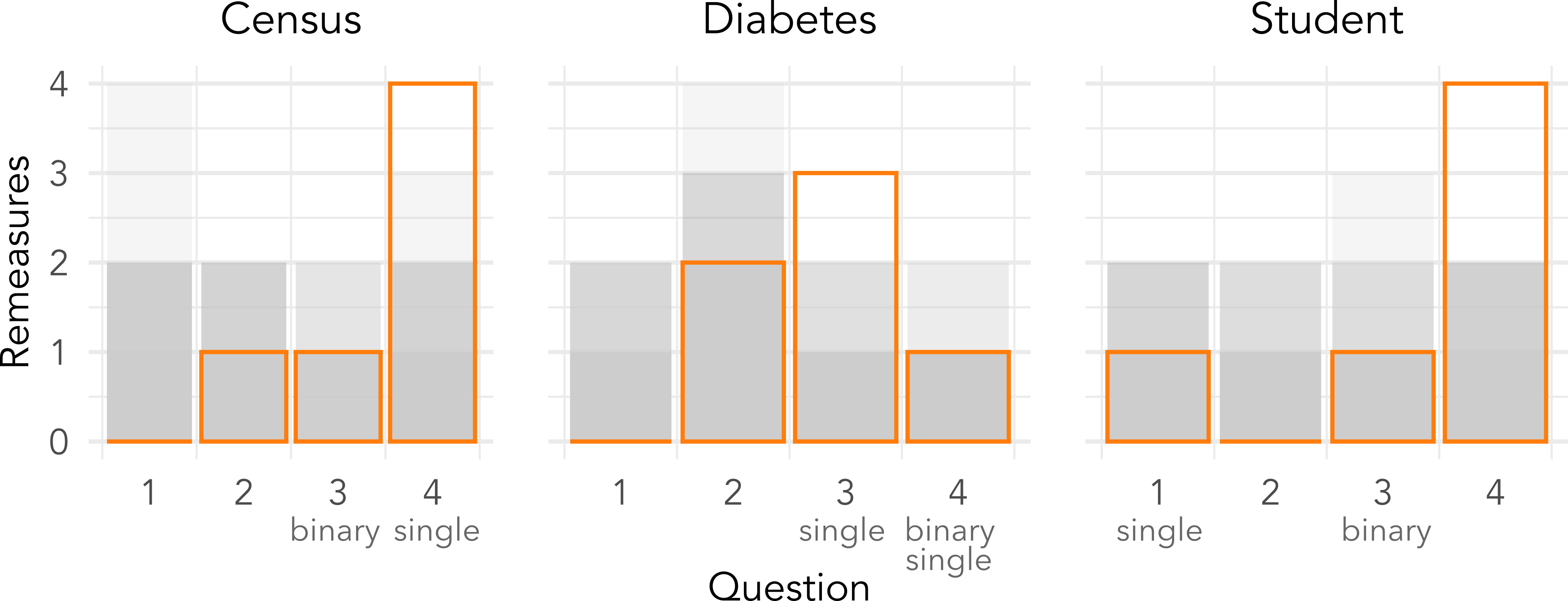}
  \caption{Remeasure allocations for each participant (gray) and the rational agent under the ex-ante allocation (orange). Questions not labeled ``binary'' are ``quantitative'' and those not labeled ``single'' are multi-value.}
  \label{fig:remeasures}
\end{figure}

%% file: 10_metareview.tex
\newpage 

\section{Meta-Review}

The following meta-review was prepared by the program committee for the 2024
IEEE Symposium on Security and Privacy (S\&P) as part of the review process as
detailed in the call for papers.

\subsection{Summary}
The paper looks at differential privacy in the context of exploratory data analysis and proposes an interactive analysis approach in which analysts first obtain estimates and then expend more of their privacy budget on follow-up queries. This approach is supported by an interactive visualization. The paper further presents a user study ($n=14$) demonstrating that analysts can use the proposed approach pretty well when compared to various models of rational agents without extensive knowledge of differential privacy.

\subsection{Scientific Contributions}
\begin{itemize}
\item Creates a New Tool to Enable Future Science
\item Provides a Valuable Step Forward in an Established Field
\end{itemize}

\subsection{Reasons for Acceptance}
\begin{enumerate}
\item The paper provides a valuable step forward in an established field. This paper presents a new paradigm for interactive Differential Privacy (DP) to be used to support Exploratory Data Analysis (EDA). DP and EDA are inherently at odds as DP is meant to protect against probing/exploration of databases to protect against differencing attacks. But, as pointed out in the paper, there are very legitimate reasons to have to do exploratory data analysis and the paper provides an interesting solution for reconciling that need with DP protections.

\item The paper creates a new tool to enable future science. The paper operationalizes the \workflow{} paradigm through an interactive visualization. They also make their code and data to model rational agents publicly available. This allows for both independent confirmation and allows the tool to be used in future research.
\end{enumerate}

\subsection{Noteworthy Concerns} 
\begin{enumerate} 
\item The study is conducted with a small sample ($n=14$) of participants. A larger sample size would be preferable to obtain more robust results, but this size is sufficient as this is exploratory work.

\item Participants were recruited from graduate student listservs and through the authors' personal networks. Participants were not asked to report their professional experience, so it is not clear whether the sample included professionals, which would be preferable as professionals might bring additional expertise to the task. However, at a minimum, these results provide insights for early career analysts with limited DP experience, and the study is designed to limit the need for domain knowledge.
\end{enumerate}

%% file: main.bbl
\begin{thebibliography}{10}
\providecommand{\url}[1]{#1}
\csname url@samestyle\endcsname
\providecommand{\newblock}{\relax}
\providecommand{\bibinfo}[2]{#2}
\providecommand{\BIBentrySTDinterwordspacing}{\spaceskip=0pt\relax}
\providecommand{\BIBentryALTinterwordstretchfactor}{4}
\providecommand{\BIBentryALTinterwordspacing}{\spaceskip=\fontdimen2\font plus
\BIBentryALTinterwordstretchfactor\fontdimen3\font minus \fontdimen4\font\relax}
\providecommand{\BIBforeignlanguage}[2]{{%
\expandafter\ifx\csname l@#1\endcsname\relax
\typeout{** WARNING: IEEEtran.bst: No hyphenation pattern has been}%
\typeout{** loaded for the language `#1'. Using the pattern for}%
\typeout{** the default language instead.}%
\else
\language=\csname l@#1\endcsname
\fi
#2}}
\providecommand{\BIBdecl}{\relax}
\BIBdecl

\bibitem{steed2022policy}
R.~Steed, T.~Liu, Z.~S. Wu, and A.~Acquisti, ``Policy impacts of statistical uncertainty and privacy,'' \emph{Science}, vol. 377, no. 6609, pp. 928--931, 2022.

\bibitem{dwork2006calibrating}
C.~Dwork, F.~McSherry, K.~Nissim, and A.~Smith, ``Calibrating noise to sensitivity in private data analysis,'' in \emph{Theory of Cryptography: Third Theory of Cryptography Conference, TCC 2006, New York, NY, USA, March 4-7, 2006. Proceedings 3}.\hskip 1em plus 0.5em minus 0.4em\relax Springer, 2006, pp. 265--284.

\bibitem{dwork2014algorithmic}
C.~Dwork and A.~Roth, ``The algorithmic foundations of differential privacy,'' \emph{Foundations and Trends{\textregistered} in Theoretical Computer Science}, vol.~9, no. 3--4, pp. 211--407, 2014.

\bibitem{sarathy2023don}
J.~Sarathy, S.~Song, A.~Haque, T.~Schlatter, and S.~Vadhan, ``Don’t look at the data! how differential privacy reconfigures the practices of data science,'' in \emph{Proceedings of the 2023 CHI Conference on Human Factors in Computing Systems}, 2023, pp. 1--19.

\bibitem{wasserman2012minimaxity}
L.~Wasserman, ``Minimaxity, statistical thinking and differential privacy,'' \emph{Journal of Privacy and Confidentiality}, vol.~4, no.~1, 2012.

\bibitem{tukey1977exploratory}
J.~W. Tukey \emph{et~al.}, \emph{Exploratory data analysis}.\hskip 1em plus 0.5em minus 0.4em\relax Reading, MA, 1977, vol.~2.

\bibitem{simon1990bounded}
H.~A. Simon, ``Bounded rationality,'' \emph{Utility and probability}, pp. 15--18, 1990.

\bibitem{nunez2020every}
S.~Nu{\~n}ez~von Voigt, M.~Pauli, J.~Reichert, and F.~Tschorsch, ``Every query counts: Analyzing the privacy loss of exploratory data analyses,'' in \emph{Data Privacy Management, Cryptocurrencies and Blockchain Technology: ESORICS 2020 International Workshops, DPM 2020 and CBT 2020, Guildford, UK, September 17--18, 2020, Revised Selected Papers 15}.\hskip 1em plus 0.5em minus 0.4em\relax Springer, 2020, pp. 258--266.

\bibitem{mckenna2018optimizing}
R.~McKenna, G.~Miklau, M.~Hay, and A.~Machanavajjhala, ``Optimizing error of high-dimensional statistical queries under differential privacy,'' \emph{arXiv preprint arXiv:1808.03537}, 2018.

\bibitem{mckenna2021hdmm}
{McKenna, Ryan and Miklau, Gerome and Hay, Michael and Machanavajjhala, Ashwin}, ``{HDMM: Optimizing error of high-dimensional statistical queries under differential privacy},'' \emph{VLDB}, 2021.

\bibitem{li2015matrix}
C.~Li, G.~Miklau, M.~Hay, A.~McGregor, and V.~Rastogi, ``The matrix mechanism: optimizing linear counting queries under differential privacy,'' \emph{The VLDB journal}, vol.~24, pp. 757--781, 2015.

\bibitem{ngong2023evaluating}
I.~C. Ngong, B.~Stenger, J.~P. Near, and Y.~Feng, ``Evaluating the usability of differential privacy tools with data practitioners,'' \emph{arXiv preprint arXiv:2309.13506}, 2023.

\bibitem{garrido2023lessons}
G.~M. Garrido, X.~Liu, F.~Matthes, and D.~Song, ``Lessons learned: Surveying the practicality of differential privacy in the industry,'' \emph{Privacy Enhancing Technologies (PETS)}, 2023.

\bibitem{gaboardi2018psi}
M.~Gaboardi, J.~Honaker, G.~King, J.~Murtagh, K.~Nissim, J.~Ullman, and S.~Vadhan, ``Psi ({$\Psi$}): a private data sharing interface,'' 2018.

\bibitem{thaker2020overlook}
P.~Thaker, M.~Budiu, P.~Gopalan, U.~Wieder, and M.~Zaharia, ``Overlook: Differentially private exploratory visualization for big data,'' \emph{Theory and Practice of Differential Privacy (TPDP)}, 2020.

\bibitem{nanayakkara2022visualizing}
P.~Nanayakkara, J.~Bater, X.~He, J.~Hullman, and J.~Rogers, ``Visualizing privacy-utility trade-offs in differentially private data releases,'' \emph{Proceedings on Privacy Enhancing Technologies}, vol. 2022, no.~2, pp. 601--618, 2022.

\bibitem{john2021decision}
M.~F.~S. John, G.~Denker, P.~Laud, K.~Martiny, A.~Pankova, and D.~Pavlovic, ``Decision support for sharing data using differential privacy,'' in \emph{2021 IEEE Symposium on Visualization for Cyber Security (VizSec)}.\hskip 1em plus 0.5em minus 0.4em\relax IEEE, 2021, pp. 26--35.

\bibitem{ge2019apex}
C.~Ge, X.~He, I.~F. Ilyas, and A.~Machanavajjhala, ``Apex: Accuracy-aware differentially private data exploration,'' in \emph{Proceedings of the 2019 International Conference on Management of Data}, 2019, pp. 177--194.

\bibitem{bittner2020understanding}
D.~M. Bittner, A.~E. Brito, M.~Ghassemi, S.~Rane, A.~D. Sarwate, and R.~N. Wright, ``Understanding privacy-utility tradeoffs in differentially private online active learning,'' \emph{Journal of Privacy and Confidentiality}, vol.~10, no.~2, 2020.

\bibitem{guo2023seeing}
Y.~Guo, F.~Liu, T.~Zhou, Z.~Cai, and N.~Xiao, ``Seeing is believing: Towards interactive visual exploration of data privacy in federated learning,'' \emph{Information Processing \& Management}, vol.~60, no.~2, p. 103162, 2023.

\bibitem{hay2016exploring}
M.~Hay, A.~Machanavajjhala, G.~Miklau, Y.~Chen, D.~Zhang, and G.~Bissias, ``Exploring privacy-accuracy tradeoffs using dpcomp,'' in \emph{Proceedings of the 2016 International Conference on Management of Data}, 2016, pp. 2101--2104.

\bibitem{hay2016principled}
M.~Hay, A.~Machanavajjhala, G.~Miklau, Y.~Chen, and D.~Zhang, ``Principled evaluation of differentially private algorithms using dpbench,'' in \emph{Proceedings of the 2016 International Conference on Management of Data}, ser. SIGMOD '16.\hskip 1em plus 0.5em minus 0.4em\relax New York, NY, USA: Association for Computing Machinery, 2016, p. 139–154.

\bibitem{zhang2018ektelo}
D.~Zhang, R.~McKenna, I.~Kotsogiannis, M.~Hay, A.~Machanavajjhala, and G.~Miklau, ``Ektelo: A framework for defining differentially-private computations,'' in \emph{Proceedings of the 2018 International Conference on Management of Data}, 2018, pp. 115--130.

\bibitem{mcsherry2009privacy}
F.~D. McSherry, ``Privacy integrated queries: an extensible platform for privacy-preserving data analysis,'' in \emph{Proceedings of the 2009 ACM SIGMOD International Conference on Management of data}, 2009, pp. 19--30.

\bibitem{wu2023rational}
Y.~Wu, Z.~Guo, M.~Mamakos, J.~Hartline, and J.~Hullman, ``The rational agent benchmark for data visualization,'' \emph{IEEE Transactions on Visualization and Computer Graphics}, 2023.

\bibitem{gneiting2007strictly}
T.~Gneiting and A.~E. Raftery, ``Strictly proper scoring rules, prediction, and estimation,'' \emph{Journal of the American statistical Association}, vol. 102, no. 477, pp. 359--378, 2007.

\bibitem{brier1950verification}
G.~W. Brier \emph{et~al.}, ``Verification of forecasts expressed in terms of probability,'' \emph{Monthly weather review}, vol.~78, no.~1, pp. 1--3, 1950.

\bibitem{NIST_ACS_data}
\BIBentryALTinterwordspacing
{Task C. and Bhagat K. and Streat D. and Simpson A. and Howarth G.S.}, ``{NIST Diverse Communities Data Excerpts},'' 2023. [Online]. Available: \url{https://doi.org/10.18434/mds2-2895}
\BIBentrySTDinterwordspacing

\bibitem{strack2014impact}
B.~Strack, J.~P. DeShazo, C.~Gennings, J.~L. Olmo, S.~Ventura, K.~J. Cios, and J.~N. Clore, ``Impact of hba1c measurement on hospital readmission rates: analysis of 70,000 clinical database patient records,'' \emph{BioMed research international}, vol. 2014, 2014.

\bibitem{GunduzFokoue:2013}
\BIBentryALTinterwordspacing
N.~Gunduz and E.~Fokoue, ``{UCI} machine learning repository,'' 2013. [Online]. Available: \url{http://archive.ics.uci.edu/ml/datasets/turkiye+student+evaluation}
\BIBentrySTDinterwordspacing

\bibitem{soll2004overconfidence}
J.~B. Soll and J.~Klayman, ``Overconfidence in interval estimates.'' \emph{Journal of Experimental Psychology: Learning, Memory, and Cognition}, vol.~30, no.~2, p. 299, 2004.

\bibitem{apple2017}
\BIBentryALTinterwordspacing
{Apple Differential Privacy Team}, ``Learning with privacy at scale,'' 2017. [Online]. Available: \url{https://docs-assets.developer.apple.com/ml-research/papers/learning-with-privacy-at-scale.pdf}
\BIBentrySTDinterwordspacing

\bibitem{erlingsson2014rappor}
{\'U}.~Erlingsson, V.~Pihur, and A.~Korolova, ``Rappor: Randomized aggregatable privacy-preserving ordinal response,'' in \emph{Proceedings of the 2014 ACM SIGSAC conference on computer and communications security}, 2014, pp. 1054--1067.

\bibitem{ding2017collecting}
B.~Ding, J.~Kulkarni, and S.~Yekhanin, ``Collecting telemetry data privately,'' \emph{Advances in Neural Information Processing Systems}, vol.~30, 2017.

\bibitem{tezapsidis2017uber}
\BIBentryALTinterwordspacing
K.~Tezapsidis, ``Uber releases open source project for differential privacy,'' 2017. [Online]. Available: \url{https://medium.com/uber-security-privacy/differential-privacy-open-source-7892c82c42b6}
\BIBentrySTDinterwordspacing

\bibitem{abowd2018us}
J.~M. Abowd, ``The us census bureau adopts differential privacy,'' in \emph{Proceedings of the 24th ACM SIGKDD International Conference on Knowledge Discovery \& Data Mining}, 2018, pp. 2867--2867.

\bibitem{zhang2016challenges}
D.~Zhang, M.~Hay, G.~Miklau, and B.~O’Connor, ``Challenges of visualizing differentially private data,'' \emph{Theory and Practice of Differential Privacy}, vol. 2016, pp. 1--3, 2016.

\bibitem{zhou2022dpviscreator}
J.~Zhou, X.~Wang, J.~K. Wong, H.~Wang, Z.~Wang, X.~Yang, X.~Yan, H.~Feng, H.~Qu, H.~Ying \emph{et~al.}, ``Dpviscreator: Incorporating pattern constraints to privacy-preserving visualizations via differential privacy,'' \emph{IEEE Transactions on Visualization and Computer Graphics}, vol.~29, no.~1, pp. 809--819, 2022.

\bibitem{panavas2023investigating}
L.~Panavas, T.~Crnovrsanin, J.~L. Adams, J.~Ullman, A.~Sargavad, M.~Tory, and C.~Dunne, ``Investigating the visual utility of differentially private scatterplots,'' \emph{IEEE Transactions on Visualization and Computer Graphics}, 2023.

\bibitem{hullman2015hypothetical}
J.~Hullman, P.~Resnick, and E.~Adar, ``Hypothetical outcome plots outperform error bars and violin plots for inferences about reliability of variable ordering,'' \emph{PloS one}, vol.~10, no.~11, p. e0142444, 2015.

\bibitem{garrido2021get}
G.~M. Garrido, J.~Near, A.~Muhammad, W.~He, R.~Matzutt, and F.~Matthes, ``Do i get the privacy i need? benchmarking utility in differential privacy libraries,'' \emph{arXiv preprint arXiv:2109.10789}, 2021.

\end{thebibliography}
